\pdfoutput=1
\documentclass[aps,prl,twocolumn,amsmath,amssymb,floatfix]{revtex4-2}

\usepackage{graphicx}
\usepackage{bm}
\usepackage{xcolor}
\usepackage[colorlinks=true,linkcolor=blue!50!black,citecolor=blue!50!black,urlcolor=blue!50!black]{hyperref}

\graphicspath{{figures/}}
\newcommand{\dcont}{d_{\mathrm{cont}}}
\newcommand{\Tenc}{T_{\mathrm{enc}}}
\newcommand{\tenc}{\tau_{\mathrm{enc}}}
\newcommand{\med}{\mathrm{med}}
\newcommand{\cond}{\,|\,k>0}
\newcommand{\avgc}[1]{{\langle #1\cond\rangle}}
\newcommand{\IAR}{I(A{:}R)}

\begin{document}

\title{Small-world networks yield tunable quantum codes at short circuit depth}

\author{Samuel Bevins}
\author{Gregory Bentsen}
\affiliation{Department of Physics, William \& Mary, Williamsburg, Virginia 23187, USA}

\date{\today}

\begin{abstract}
Monitored random quantum circuits naturally generate quantum error-correcting codes. We generate quantum codes with tunable properties by implementing random Clifford circuits on Watts--Strogatz `small-world'
networks and tailor code properties via three knobs: nonlocal graph rewiring, measurement rate, and circuit depth. On fully rewired networks, circuits of
logarithmic depth produce codes with a finite rate and an extensive mean contiguous distance with failure probability substantially lower than codes produced from local circuits. These small-world codes form at a circuit depth growing as $\ln n$ compared to polynomial depths required for codes produced from local networks.
Rewiring also moves the purification transition to higher measurement rates and changes its critical exponents. Connectivity is thus a
knob for designing codes.
\end{abstract}

\maketitle

\begin{figure}[t]
  \centering
  \includegraphics[width=\columnwidth]{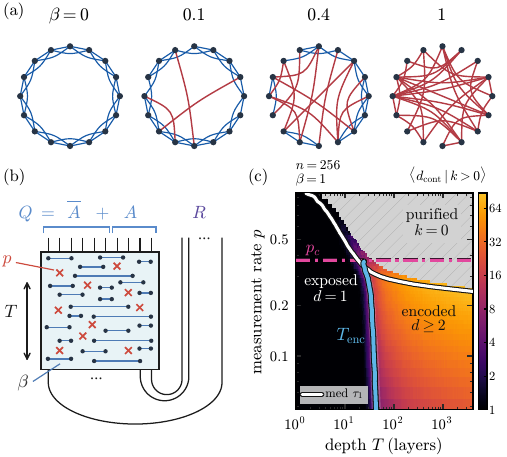}
  \caption{Tunable small-world monitored circuits. (a)~Watts--Strogatz graphs $W(16,\beta)$, ring bonds in blue, rewired bonds in red. (b)~Monitored circuit: the system $Q=\overline{A}\cup A$ in Bell
  pairs with the reference $R$; $T$ layers of gates on graph bonds and measurements at rate $p$ (crosses).
  (c)~Phase diagram at $n=256$ on fully rewired graphs ($\beta=1$), to depth $T=16n$. Color: $\avgc{\dcont}$ where at least 10\% of the circuits hold a code; hatched gray: fewer do.
  Cyan: the encoding time $\Tenc$; legend: the median of the first layer $\tau_1$ with $k\le1$; pink dash-dotted: the
  critical rate $p_c=0.376$ of Fig.~\ref{fig:fss}.}
  \label{fig:model}
\end{figure}

\textit{Introduction.}---Quantum error-correcting codes with a finite rate and an extensive distance could greatly
reduce the overhead of fault-tolerant quantum computation. Quantum low-density parity-check codes achieve
both~\cite{panteleev2022}, but they are highly structured and need nonlocal connectivity, since local codes in two
dimensions obey stringent rate--distance bounds~\cite{bravyi2010}. Hardware is beginning to offer such
nonlocal connectivity~\cite{bluvstein2024,bluvstein2026}, and codes~\cite{xu2024,zhao2026} and
architectures~\cite{zhou2025isca} are being co-designed to leverage these capabilities. However, adapting these codes to different hardware geometries and constraints can require substantial redesign. This motivates a complementary approach: a flexible procedure that generates quantum codes directly from the available interaction network.

Monitored random quantum circuits, which interleave entangling gates with projective measurements, suggest such a procedure. These circuits
generate quantum error-correcting codes~\cite{choi2020} where the measurement-induced phase transition~\cite{li2018,skinner2019,chan2019,li2019,bao2020,jian2020,zabalo2020,gullans2020prl} marks a
transition in their capacity to protect quantum information~\cite{choi2020,gullans2020prx,fan2021self,li2021,fisher2023}. Purely unitary random circuits already provide
polylogarithmic-depth encoders for good quantum codes~\cite{brown2013,gullans2021,anand2026good}, achieving
decoupling~\cite{hayden2007} and approaching erasure-channel capacity even with local interactions in any spatial
dimension~\cite{gullans2021}. These codes also admit efficient Pauli-noise decoding~\cite{darmawan2024low} and
fault-tolerant state preparation~\cite{nelson2025fault}. However, unitary evolution preserves the chosen input code
rate, and unitary scrambling tends to generate high-weight stabilizer checks, making these codes unwieldy for realistic experiments.

Monitoring and nonlocal interactions
provide additional control. Projective measurements naturally tune the code rate and continually introduce low-weight
stabilizers~\cite{gullans2020prx,li2021}. By removing exposed logical information, they also provide a stochastic
analogue of the targeted measurement-based expurgation proposed in Ref.~\cite{gullans2021}. In addition, while previous work has focused largely on local lattice interactions, leveraging nonlocal connectivity offers further opportunities.
Nonlocal networks rapidly delocalize quantum information, where sparse interaction graphs with a finite Cheeger constant and a nonvanishing Laplacian gap $\lambda_2$~\cite{fiedler1973}
can globally scramble information in logarithmic time~\cite{sekino2008,bentsen2019}.
Nonlocal interactions also reshape measurement-induced criticality~\cite{nahum2021,bentsen2021,block2022,sharma2022measurement,feng2023}:
sparse nonlocal circuits withstand higher measurement rates than local ones, and strongly nonlocal circuits
yield nearly extensive contiguous code distances~\cite{hashizume2022}. Related connections between nonlocal
dynamics and coding arise in the low-energy states of all-to-all interacting models~\cite{bentsen2024}.
Reconfigurable Rydberg arrays make such geometries experimentally relevant by allowing atoms to be spatially rearranged
between gates~\cite{bluvstein2024,bluvstein2026}. Together, these results motivate monitored random circuits on
nonlocal interaction networks as a flexible framework for exploring code rates, distances, and stabilizer
weights within hardware connectivity constraints.

Here we use monitored random Clifford circuits on Watts--Strogatz small-world networks~\cite{watts1998,olfatisaber2005} as a structured search for
quantum codes, using the measurement rate $p$, the rewiring probability $\beta$, and the circuit depth $T$ as tunable knobs (Fig.~\ref{fig:model}). The
small-world regime, in which the mean path length grows as $\ln n$ rather than $n$, begins once a few of the $2n$
bonds are rewired, at $\beta\sim1/n$~\cite{newman1999,barrat2000,SM}. On fully rewired networks ($\beta=1$), circuits of logarithmic
depth generate codes whose mean contiguous distance is extensive at finite rate, in contrast to circuits built on a 1D ring ($\beta=0$).
These small-world codes also fail far less often under random erasure and form faster: their encoding time grows as $\ln n$,
compared to polynomial time on the 1D ring.
Extending the statistical mechanics picture of random circuits to small-world networks~\cite{li2021,li2023dpre}, we propose a domain-wall mechanism for the robustness of these codes where an extensive difference in domain-wall tension between pinned and depinned domain walls allows for large protected subregions that scale with system size.

\begin{figure}[t]
  \centering
  \includegraphics[width=\columnwidth]{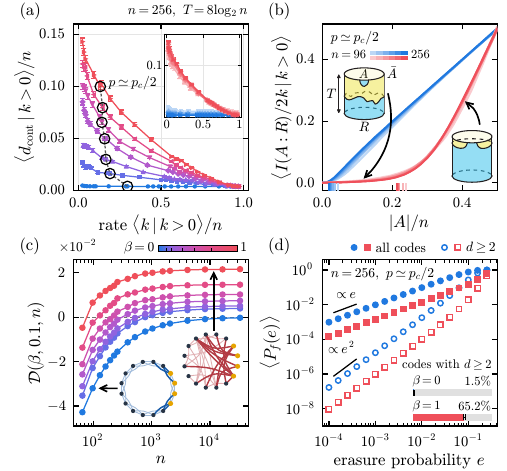}
  \caption{Small-world code properties.
  (a)~Rate--distance trade-off, one point per $p$, for $\beta$ from 0 (blue) to 1 (red); circles: $p\simeq p_c/2$;
  inset: $\beta=0$ and 1, $n=96$--256.
  (b)~Share of the encoded information held by the first $|A|$ qubits; ticks: mean decoupling size $A^*$. Cartoons: the
  circuit between $R$ (bottom) and $A\cup\bar A$ (top), its domain wall bound to the top ($A$ exposed) or free in the
  bulk ($A$ protected).
  (c)~Tension difference $\mathcal D(\beta,0.1,n)$~\cite{SM}; insets: a pinned arc $A$ (gold) and its bonds to the
  rest of the graph at $\beta=0$ and 1.
  (d)~Mean failure probability $\langle P_f(e)\rangle$ under i.i.d.\ erasure, for all codes and those with $d\ge2$;
  bars: the share with $d\ge2$. Error bars: one standard error over graphs.}
  \label{fig:codes}
\end{figure}

\textit{Model and observables.}---We construct monitored random Clifford circuits on Watts--Strogatz networks
$W(n,\beta)$~\cite{watts1998}, tuning interaction nonlocality through the rewiring probability $\beta$,
monitoring through the measurement rate $p$, and encoding time through the circuit depth $T$. To construct
$W(n,\beta)$, we begin with a degree-four ring of $n$ vertices, each connected to its two nearest neighbors on
either side, and independently rewire each of its $2n$ edges with probability $\beta$. To rewire the graph, we visit each edge $\{i,j\}$ in turn, retain vertex $i$, and replace $j$ by a uniformly
chosen vertex distinct from $i$ and not already adjacent to it~\cite{olfatisaber2005}. This preserves the total number of edges while excluding self-loops and duplicates. Figure~\ref{fig:model}(a) illustrates the progression from the local 1D ring ($\beta=0$)
to fully rewired graphs ($\beta=1$).

Each vertex in the graph $W(n,\beta)$ hosts one system qubit, and graph edges specify the allowed two-qubit interactions
[Fig.~\ref{fig:model}(b)]. At each circuit layer we apply $n/2$ random gates in sequence, independently sampling a uniformly random
two-qubit Clifford gate acting on an edge uniformly drawn from $W(n,\beta)$. Each system qubit is then independently measured in the $Z$ basis with
probability $p$. We repeat this procedure for $T$ layers on the same graph and thereby obtain a stabilizer tableau for each Clifford circuit~\cite{aaronson2004,sparsegf2}. Gates spread information, while
measurements remove it and define a code conditioned on the measurement record. This procedure yields an ensemble of codes; throughout, angled brackets $\langle \ldots \rangle$ denote averages over the ensemble at fixed $(n,\beta,p,T)$.

To characterize the resulting code, we initially maximally entangle the system qubits with a reference $R$. For each trajectory, the second R\'enyi
entropy $S_R=-\log_2\operatorname{Tr}\rho_R^2$ counts the surviving logical qubits, $k=S_R$, giving the code rate $k/n$.
The second R\'enyi mutual information $\IAR=S_A+S_R-S_{AR}$ quantifies how much logical information is exposed by an output subregion $A$~\footnote{Stabilizer states generated by Clifford circuit dynamics have flat entanglement spectra, so these R\'enyi entropies equal their von Neumann counterparts.}.
A region is
\emph{protected} when $\IAR=0$: it contains no logical information, and erasure of $A$ is correctable using the
remaining output and measurement record~\cite{schumacher1996}. It is \emph{exposed} when $\IAR>0$, so errors
supported there can compromise logical information.
The competition between gates and measurements determines both how much information survives and how well it is
protected.

To quantify code robustness, one typically computes the distance \(d\), the size of the smallest exposed subregion.
Because finding it generally requires an
exponential search, our primary diagnostic is the contiguous distance $\dcont$, the length of the shortest
exposed arc along the original ring~\cite{li2021}, which upper-bounds the true distance $\dcont\ge d$. Although natural for the local geometry at $\beta = 0$, it remains a useful
common probe across all $\beta$. In the fully rewired network $\beta = 1$ contiguity loses its geometrical meaning, but in that case any choice of subregion $A$ will serve since all vertices are statistically equivalent.
We also test protection on arbitrary subsets through random erasure below.

\begin{figure}[t]
  \centering
  \includegraphics[width=\columnwidth]{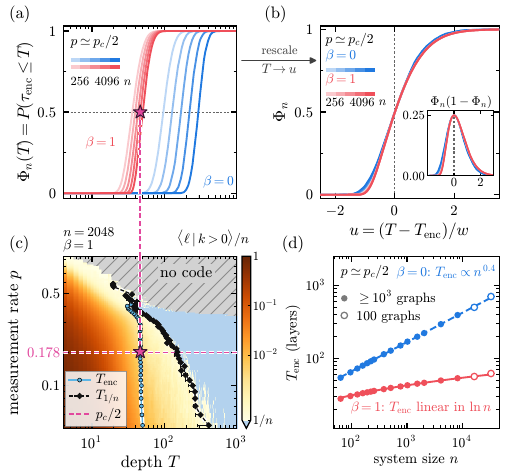}
  \caption{Short-depth coding transition. Colors as in Fig.~\ref{fig:codes}. (a)~Encoding
  fronts $\Phi_n(T)=P(\tenc\le T)$, the share of circuits encoded by depth $T$, at $p\simeq p_c/2$, on the depth axis
  of (c); dashed: $\Phi_n=1/2$, reached at $\Tenc$. (b)~The same fronts against $u=(T-\Tenc)/w$ (arrow), with $w$
  their interquartile width; inset: their variance $\Phi_n(1-\Phi_n)$ over circuits. (c)~Mean fraction $\avgc{\ell}/n$ of exposed qubits at $n=2048$ and $\beta=1$, with
  $\Tenc$ and $T_{1/n}$, past which $\avgc{\ell}<1$ (light blue); dashed: $p\simeq p_c/2$ and the $\Tenc$ of
  (a) at this $n$ and $\beta$, which meet on the $\Tenc$ front (stars). (d)~$\Tenc$ against $n$ at the rates of
  (b); lines: fits over $n\ge512$, $\propto n^{0.402}$ on the ring and $a+5.01\ln n$ at $\beta=1$.
  Bands in (a) and error bars: 16--84\% over graph resamples.}
  \label{fig:enc}
\end{figure}

\begin{figure}[t]
  \centering
  \includegraphics[width=\columnwidth]{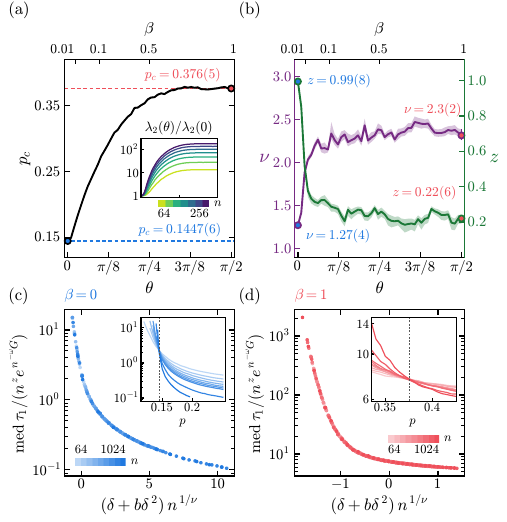}
  \caption{Measurement-induced purification from the ring to fully rewired graphs, from Eq.~(\ref{eq:fss})
  fitted at each angle to $y=\ln\med\,\tau_1$ over $n=64$--1024; top axes: $\beta=\sin^2\theta$; blue and red: the ring and
  $\beta=1$; their values, with statistical and systematic errors combined, as dashes in (a) and beside the end points in (b). (a)~The critical rate $p_c$. Inset: the mean algebraic connectivity $\lambda_2$ relative to the ring's, for
  $n=64$--256 (key). (b)~The correlation-length exponent $\nu$ (plum, left) and the dynamical exponent
  $z$ (green, right). (c),~(d)~Collapses
  $\med\,\tau_1/\bigl(n^{z}e^{n^{-\omega}G(x)}\bigr)=e^{F(x)}$ against $x$ at $\beta=0$ and 1, lighter for smaller $n$ (keys);
  insets: the same against $p$, crossing at $p_c$ (dashed). Bands in (a) and (b) (in (a) narrower than the line):
  16--84\% over 2000 graph resamples, statistical only.}
  \label{fig:fss}
\end{figure}

Our small-world random circuit codes yield a `phase diagram' [Figure~\ref{fig:model}(c)] which shows three distinct regimes in the $(T,p)$ plane at $\beta=1$ and $n = 256$, diagnosed by the rate $k$ and mean contiguous
distance. At large measurement rates $p>p_c$, measurements lead to a \emph{purified} regime with $k=0$ (hatched gray). Below $p_c$,
shallow circuits ($T<\Tenc$) form an \emph{exposed} regime with vulnerable single qubits; deeper circuits ($T>\Tenc$) enter an
\emph{encoded} regime with large contiguous distance. 
In the following sections we systematically characterize different parts of this `phase diagram.' In particular, we examine the code properties of the encoded regime in Fig.~\ref{fig:codes}, the encoding-depth boundary in
Fig.~\ref{fig:enc}, and the purification transition in Fig.~\ref{fig:fss}.

\textit{Code properties.}---Fully rewired graphs generate codes with a favorable rate--distance trade-off
already at logarithmic depth, $T=8\log_2n$. Figure~\ref{fig:codes}(a) plots $\avgc{\dcont}/n$ against code rate
$\avgc{k}/n$, with each point corresponding to a measurement rate $p$, and error bars obtained from resampling~\cite{efron1979}. Code quality improves dramatically as $\beta$ increases from zero (blue) to
one (red), with codes protecting larger contiguous regions at each fixed rate. For $\beta=1$, the
curves collapse across $n=96$--256 (inset, red), consistent with an extensive mean contiguous distance at finite
rate over these sizes. By contrast, the rate-distance curves for the 1D ring (inset, blue) instead deteriorate with increasing $n$. We emphasize that these results are
ensemble averages; individual codes can still have $d=1$~\cite{SM}.

Vanishing mutual information, $\IAR=0$, means that erasing $A$ reveals no logical information and is
correctable. Figure~\ref{fig:codes}(b) shows this protection extending to large subregions on fully rewired
graphs (red). Growing a contiguous arc from a fixed endpoint, the mean decoupling subregion size $A^*$---the largest arc with
$\IAR=0$---extends out to extensive sizes $A^* \sim 0.2n$. By contrast, on the 1D ring $\beta = 0$ (blue), small subregions expose information almost immediately. The decoupling subregion size $A^*$ can be understood through a domain-wall depinning mechanism in the
Ising description of monitored circuits~\cite{bao2020,jian2020,shtanko2020classical,fan2021self,li2021,li2023dpre,li2024statistical,zhou2019emergent}. The inset cartoons in Figure~\ref{fig:codes}(b) illustrate this mechanism, where an exposed
region (right) shares a blue domain with the reference $R$, while a protected region (left) is separated from $R$ by a bulk
domain wall. Free to fluctuate through the circuit interior, the depinned bulk domain wall gains configurational entropy
that can outweigh the energetic cost $2\sigma|A|$, where $\sigma$ is the domain-wall tension. Below $A^*$,
this entropy favors decoupling; larger regions favor pinning and expose logical information.

On the 1D ring, the additional bulk entropy comes primarily from translating
the wall in time, giving $2\sigma A^*\sim\ln T$~\cite{li2021}. Here, pinning constrains only the interface endpoints, leaving the pinned and bulk walls with the
same effective tension in the thermodynamic limit. By contrast, for fully rewired graphs we conjecture that
long-range connectivity makes the entropy density sensitive to pinning. In this case, pinning the domain wall reduces the
configurational entropy by an extensive amount, reducing the bulk tension $\tilde\sigma<\sigma$ below the effective tension $\sigma$ of the pinned wall, yielding an extensive decoupling subregion size
$2\sigma A^*\sim(\sigma-\tilde\sigma)n$. Thus an extensive decoupling size arises from the extensive \emph{loss}
of entropy upon pinning. A
phenomenological Gaussian interface model~\cite{SM} supports this picture. The graph Laplacian spectrum
governs the bulk domain-wall height fluctuations, while the grounded spectrum describes the
pinned interface. Comparing these spectra numerically yields a tension difference $\mathcal D(\beta,a,n) = \sigma - \tilde \sigma$ that vanishes
in the thermodynamic limit on the ring but approaches a nonzero constant on fully rewired graphs [Fig.~\ref{fig:codes}(c)].

Our resuts so far are probed by the ensemble-average contiguous distance, but this measure can miss rare exposed regions even
when most regions of the same size are protected. To address this weakness, we study information exposure on arbitrary subsets using a random erasure model. We compute $P_f(e)$, the probability that independently erasing
each qubit with probability $e$ erases logical information~\cite{grassl1997,bennett1997}. This measures both the
size and prevalence of exposed regions and is relevant to neutral-atom devices that convert dominant errors to
erasures~\cite{wu2022}. Figure~\ref{fig:codes}(d) demonstrates that the code ensemble generated from circuits on the 1D ring (blue) fail substantially more often than those generated by the fully rewired network (red) for $n=256$ at $T=64$ and $p\simeq p_c/2$; the
erasure estimator is described in the Supplemental Material~\cite{SM}. Both ensemble failure rates are
linear in $e$ at small erasure probability, reflecting codes with exposed single qubits. We can substantially improve this performance by post-selecting codes from the full ensemble. Restricting to codes with $d\ge2$
improves the failure rate to $\sim e^2$, although fewer than 2\% of 1D ring codes survive this restriction compared to over 65\% of fully rewired codes. Rewiring thus improves random-erasure robustness even when the worst-case distance is small.

\textit{Encoding at short depth.}---We have demonstrated the robustness of codes generated on small-world
networks and proposed a domain-wall mechanism for their protection. Their practical value also depends on
how rapidly this protection develops as a function of circuit depth $T$. We now show that small-world circuits encode information at circuit depths growing
only logarithmically with system size.
To probe this timescale, we define the per-circuit encoding depth $\tenc$ as the first circuit layer containing no exposed single qubits, $\ell=0$ (assuming logical information
survives, $k>0$). Figure~\ref{fig:enc}(a) plots $\Phi_n(T)=P(\tenc\le T)$, the fraction of circuits with zero single exposed qubits by depth $T$. The curves rise from zero to one at a characteristic \emph{encoding depth} $\Tenc$ at which
$\Phi_n(\Tenc)=1/2$. For small-world networks (red), the curves cluster at short depths, around 50 layers across
$n=256-4096$. By contrast, for the 1D local ring (blue), the crossover shifts strongly toward larger depths as $n$ increases,
requiring hundreds of circuit layers at the largest sizes shown.

Rescaling depth as $u=(T-\Tenc)/w$, with $w$ the interquartile width of the encoding-depth distribution, produces
a strong collapse for both geometries [Fig.~\ref{fig:enc}(b)]. Ensemble fluctuations are strong near $T=\Tenc$, with individual circuits showing wide variation in whether or not they have encoded. The inset quantifies this variation using the variance
$\Phi_n(1-\Phi_n)$ of the binary indicator that a circuit has encoded by depth $T$. It peaks at $\Phi_n=1/2$,
providing a complementary way to locate $\Tenc$.
The crossover width $w$ grows with system size, but its ratio to $\Tenc$
slowly decreases~\cite{SM}. We find no evidence of a sharp phase transition in circuit depth and therefore conservatively refer to the encoding depth as a crossover.

The number of exposed single qubits $\ell$ reveals a longer tail of vulnerable circuits as depth increases. Figure~\ref{fig:enc}(c) shows
$\avgc{\ell}/n$ on a logarithmic color scale versus $T$ and $p$, for $\beta=1$ and $n=2048$ and shows four distinct regimes. Above $p_c$, measurements purify the system and no code survives (hatched gray). Below $p_c$, shallow
circuits retain many exposed qubits $\ell > 1$. The blue dots mark the encoding depth $\Tenc$, where half the circuits have reached $\ell=0$;
this boundary stays near 50 layers over a broad range of measurement rates. Beyond this, a substantial mean exposure
can persist because the remaining vulnerable circuits contribute disproportionately. Only at larger depths
$T_{1/n}$ (black dots) does $\avgc{\ell}$ fall below $1$
(light blue). The distinction between $\Tenc$ and $T_{1/n}$ separates typical encoding from suppression of the ensemble's residual
exposure, which takes longer as $p$ decreases.

Crucially, the encoding depth $\Tenc$ scales logarithmically in the system size for small-world circuits, in contrast to the polynomial depth required for 1D local circuits. Plotting $\Tenc$ directly against $n$ makes the scaling contrast explicit [Fig.~\ref{fig:enc}(d)]. At
$p\simeq p_c/2$, fits over system sizes $n = 2^9, \ldots, 2^{15}$ yield $\Tenc\propto n^{0.402}$ for 1D local circuits, whereas fully rewired
small-world circuits are consistent with $\Tenc=a+5.01\ln n$~\cite{SM}.
It is also interesting that very little rewiring is necessary to achieve logarithmic encoding depth: the onset of this acceleration in encoding already occurs when only a few ring bonds are rewired, and most of
the reduction is achieved by $\beta\approx0.1$~\cite{SM}. Together with the code properties in
Fig.~\ref{fig:codes}, these results show that fully rewired small-world networks support robust quantum codes
whose typical encoding requires only logarithmic circuit depth.

\textit{Purification transition.}---Rewiring also changes the purification transition at which the circuit loses its
capacity to retain quantum information. We characterize this transition using $\tau_1$, the first circuit layer
at which the number of logical qubits falls to $k\le1$. Below the critical measurement rate $p < p_c$ the typical purification time grows exponentially with
system size; by contrast, above $p_c$ purification occurs at logarithmic circuit depths~\cite{gullans2020prx}. To locate the transition throughout the range $\beta = 0,\ldots,1$,
we fit $y=\ln\med\,\tau_1$ for $n=64$--1024 to the finite-size scaling ansatz~\cite{slevin1999}
\begin{equation}
  y=z\ln n+F(x)+n^{-\omega}G(x),\qquad x=(\delta+b\,\delta^2)\,n^{1/\nu},
  \label{eq:fss}
\end{equation}
where $\delta=p-p_c$, $z$ is the dynamical exponent, and $\nu$ is the correlation length critical exponent.
Whereas the leading scaling function $F$ controls critical scaling in the thermodynamic limit, the correction function $G$ is important (especially in the nonlocal regime) to obtain good scaling collapse at finite system size. The parameter $\omega>0$ sets
the decay of finite-size corrections, and $b$ accounts for a nonlinear dependence of the scaling variable on
measurement rate. We fit each rewiring probability $\beta$ separately, with statistical and systematic uncertainties
estimated as described in the Supplemental Material~\cite{SM}. Our finite-size scaling results are presented in Fig.~\ref{fig:fss}, where the angle $\theta$ parameterizes the rewiring probability $\beta = \sin^2 \theta$ with respect to the
Fisher--Rao metric~\cite{rao1945,wootters1981,amari2000}.

The local 1D ring provides a benchmark for this analysis: our fits yield critical exponents $z = 0.99(8)$ and $\nu\simeq1.27(4)$, consistent with previous
studies of local monitored circuits in one dimension~\cite{li2018,gullans2020prx,li2021conformal}. Rewiring raises
the critical measurement rate $p_c$ smoothly as a function of $\beta$, increasing the measurement rate the code can withstand by more than a factor of two between
$\beta=0$ and 1 [Fig.~\ref{fig:fss}(a)]. The fitted exponents change much more rapidly as we move away from the local $\beta = 0$ limit
[Fig.~\ref{fig:fss}(b)]. Figures~\ref{fig:fss}(c) and (d) illustrate successful scaling collapse for the local 2D ring (blue) and fully rewired
graphs (red). In contrast to the smooth variation in $p_c$, the critical exponents change abruptly as we increase the nonlocality of the network, with the dynamical exponent dropping to $z \approx 0.22$ and the correlation length exponent increasing to $\nu \approx 2.3$. It would be interesting in future work to understand whether these changes are continous in $\beta$ or whether there is a sharp transition in the universality class~\cite{block2022,sharma2022measurement}. Further, whether this nonlocal critical behavior connects to the unconventional
critical scaling found on dynamical quantum trees remains
an interesting open question~\cite{nahum2021,feng2023}.

\textit{Conclusion and outlook.}---Our results demonstrate monitored random quantum circuits with nonlocal interactions provide a flexible framework for generating robust quantum error-correcting codes at short circuit depth. On fully rewired graphs, our small-world circuits encode quantum information in a
depth that grows only as $\ln n$, compared to polynomial circuit depths required on the 1D ring. Moreover, at this short depth our circuits produce codes with finite rate and extensive mean contiguous distance that fail far less often under random erasure than codes produced from local circuits. The nonlocal interactions also increase the critical measurement rate $p_c$ of the measurement-induced purification transition and modify its critical exponents.

The robustness and tunability of these small-world codes establish clear directions for future work. Foremost is to systematically study and minimize the weights of the stabilizer checks to improve the feasibility of syndrome extraction on realistic hardware. Further, it will be important to study decoding algorithms for these codes with an eye toward using tunable nonlocality and measurements to improve real-time decoding efficiency. Moreover, the results presented here are largely focused on ensemble averages; a particularly promising direction is to post-select especially good codes from this large pool of candidates to optimize for high rate and distance while minimizing stabilizer check weights. Finally, we hope to use our framework to codesign codes for particular hardware platforms where the interaction network and other tunable parameters are chosen to optimize for a particular device's constraints. These directions are natural targets for such optimization, including by machine learning.

\begin{acknowledgments}
\textit{Acknowledgments.}---This work was supported by funds provided by the Department of Physics and the Dean's Office at William \& Mary.
\end{acknowledgments}

\textit{Data availability.}---The data presented in this study can be reproduced bit for bit: each graph,
gate, and measurement is drawn from pseudorandom streams fixed by recorded master seeds~\cite{SM}. The code that
generates, analyzes, and plots the data is openly available in the folder \texttt{papers/SmallWorldMIPT} of the
SparseGF2 repository~\cite{sparsegf2}.

\textit{AI use disclosure.}---The authors used Claude Opus 5.5 (Anthropic, cloud), Qwen3.8-flash-next (Alibaba,
local), and ChatGPT-6.1 Sol (OpenAI, cloud) to write code, to prepare preliminary drafts of the text, and to assist with the literature review. The authors
wrote the final draft and verified all code, results, and references.

\nocite{newman1999,olfatisaber2005,fattal2004}
\bibliography{refs,sm_refs}

\clearpage
\onecolumngrid
\setcounter{equation}{0}
\setcounter{figure}{0}
\setcounter{table}{0}
\setcounter{section}{0}
\setcounter{secnumdepth}{2}
\renewcommand{\theequation}{S\arabic{equation}}
\renewcommand{\thefigure}{S\arabic{figure}}
\renewcommand{\thetable}{S\arabic{table}}
\renewcommand{\thesection}{S\arabic{section}}
\renewcommand{\thesubsection}{\thesection.\Alph{subsection}}
\makeatletter\renewcommand{\p@subsection}{}\makeatother
\renewcommand{\theHequation}{S\arabic{equation}}
\renewcommand{\theHfigure}{S\arabic{figure}}
\renewcommand{\theHtable}{S\arabic{table}}
\renewcommand{\theHsection}{S\arabic{section}}
\renewcommand{\theHsubsection}{\theHsection.\Alph{subsection}}
\renewcommand{\topfraction}{0.9}
\renewcommand{\bottomfraction}{0.8}
\renewcommand{\textfraction}{0.07}
\renewcommand{\floatpagefraction}{0.8}
\setcounter{topnumber}{3}
\setcounter{bottomnumber}{2}
\setcounter{totalnumber}{5}

\begin{center}
{\large\bfseries Supplemental Material for\\[2pt] ``Small-world networks yield tunable quantum codes at short circuit depth''}
\end{center}
\vspace{6pt}

This Supplemental Material defines the graphs (Sec.~\ref{sec:graphs}), the circuits and the simulations
(Sec.~\ref{sec:sim}), and the observables (Sec.~\ref{sec:obs}), and gives the analyses behind the figures of the main
text (Secs.~\ref{sec:phase}--\ref{sec:exp}). Equations, figures, and sections without the prefix S are those of the
main text.

\section{Small-world graphs}\label{sec:graphs}

\subsection{The graph ensemble}\label{sec:graphs-law}

The qubits are $V=\{0,\ldots,n-1\}$, with arithmetic modulo $n$. The ring $C(n,2)$ bonds each qubit to its nearest and
next-nearest neighbors,
\begin{equation}
E_0=\bigl\{\{i,i+r\}:\ i\in V,\ r=1,2\bigr\},\qquad |E_0|=2n,
\label{eq:ring}
\end{equation}
so every degree is 4. Each bond $\{i,i+r\}$ is visited once, as the slot $(i,i+r)$, in the order
\begin{equation}
(0,1),\ (0,2),\ (1,2),\ (1,3),\ \ldots,\ (n-1,0),\ (n-1,1).
\label{eq:order}
\end{equation}
At its visit, a slot is rewired with probability $\beta$: the bond keeps $i$, and its other end moves to
\begin{equation}
v\sim\mathrm{Unif}(J_i),\qquad J_i=V\setminus\bigl(N(i)\cup\{i\}\bigr),
\label{eq:J}
\end{equation}
a qubit not then bonded to $i$ ($N(i)$ is the current neighborhood of $i$; nothing changes if $J_i$ is empty). The graph
after the last visit is $W(n,\beta)$: the ring at $\beta=0$, fully rewired at $\beta=1$. It is simple and has exactly
the $2n$ bonds of the ring, and the number $S$ of rewiring attempts is $\mathrm{Bin}(2n,\beta)$. Since the
endpoints do not depend on $\beta$ once the attempts are fixed, the law of the ensemble is
\begin{equation}
\Pr\bigl[W(n,\beta)=g,\ S=m\bigr]=\binom{2n}{m}\beta^m(1-\beta)^{2n-m}\,Q_{n,m}(g),
\label{eq:decomp}
\end{equation}
with $Q_{n,m}$ the law of the graph when a uniformly random set of $m$ slots is rewired. Every graph we sample is connected.

\begin{figure}[tbp]
\centering
\includegraphics[width=\textwidth]{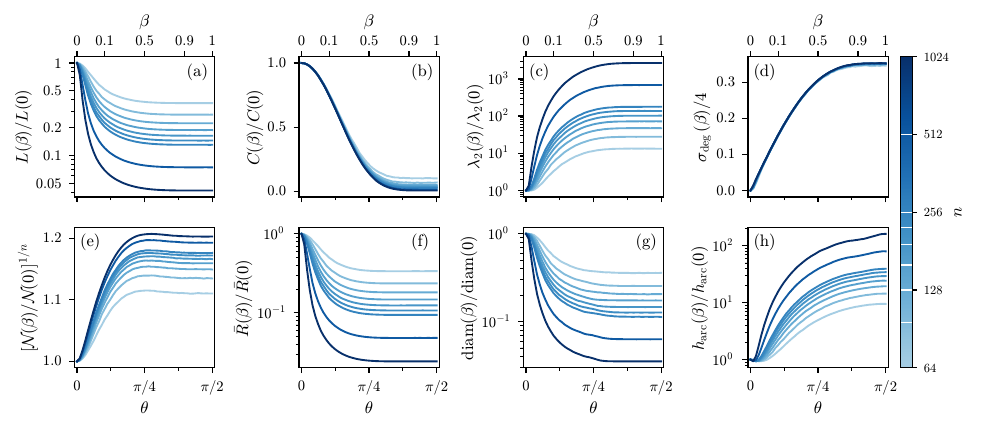}
\caption{Graph invariants of the simulated graphs against the rewiring angle $\theta$ (bottom; $\beta$ on top), each
relative to the ring ($\beta=0$) except (d): (a)~the mean path length $L$, (b)~the clustering coefficient $C$, (c)~the
algebraic connectivity $\lambda_2$, (d)~the standard deviation $\sigma_{\deg}$ of the qubit degrees, relative to the
mean degree, 4, (e)~the number of spanning trees $\mathcal N$, per qubit [Eq.~(\ref{eq:trees})], (f)~the mean effective
resistance $\bar R$ between two qubits [Eq.~(\ref{eq:resist})], (g)~the diameter $\mathrm{diam}$, and (h)~the arc expansion
$h_{\mathrm{arc}}$ [Eq.~(\ref{eq:harc})]. Each line joins the means over the 1000 graphs simulated at each of the 50
angles $\theta_j$, at one of the nine sizes $n=64$--1024 (color, logarithmic in $n$), with the mean of $\ln\mathcal N$ in (e);
the bands, standard errors over graphs, are mostly narrower than the lines.
}
\label{fig:graphs}
\end{figure}

Figure~\ref{fig:graphs} follows invariants of the simulated graphs as rewiring turns the ring into a sparse random
graph; $d_{ij}$ is the graph distance between qubits $i$ and $j$. Watts and Strogatz rewired the bonds of a ring at
random: a few shortcuts bring the mean path length $L$, the mean of $d_{ij}$ over pairs of qubits, close to that of a
random graph while the clustering stays close to the ring's~\cite{watts1998}. The path length crosses over to growing as
$\ln n$ once the ring carries of order one rewired bond, that is, at
$\beta\sim1/n$~\cite{barthelemy1999,newman1999rg,newman1999,newman2000}.

The graph Laplacian $\mathcal L=K-W$, with $W$ the adjacency matrix and $K$ the diagonal matrix of degrees, has
eigenvalues $0=\lambda_1\le\lambda_2\le\cdots\le\lambda_n$; the algebraic connectivity $\lambda_2$ is positive exactly
when the graph is connected~\cite{fiedler1973}. At $\beta=1$ it still falls slowly with $n$ over our sizes.

Rewiring keeps the $2n$ bonds, and so the mean degree 4, but spreads the degrees~\cite{barrat2000}
[Fig.~\ref{fig:graphs}(d)]. Each gate acts on a uniformly random bond (Sec.~\ref{sec:sim}), so a qubit takes part in
gates in proportion to its degree.

The number of spanning trees~\cite{kirchhoff1847} and, with every bond a unit resistor, the mean effective resistance
between two qubits~\cite{klein1993,gutman1996} [Fig.~\ref{fig:graphs}(e), (f)],
\begin{equation}
\mathcal N=\frac1n\prod_{k=2}^n\lambda_k,
\label{eq:trees}
\end{equation}
\begin{equation}
\bar R=\frac{2}{n(n-1)}\sum_{i<j}R_{ij}=\frac{2}{n-1}\sum_{k=2}^n\frac1{\lambda_k},
\label{eq:resist}
\end{equation}
are set by the geometric and harmonic means of the nonzero eigenvalues: $\ln\mathcal N$ weights the whole spectrum
alike, $\bar R$ its low end, which sets how fast a random walk joins two qubits~\cite{chandra1996}.

The diameter $\mathrm{diam}$ is the largest $d_{ij}$ [Fig.~\ref{fig:graphs}(g)]; scrambling in a time logarithmic in $n$
requires $\mathrm{diam}\lesssim\ln n$~\cite{bentsen2019}. On the ring, $\mathrm{diam}=n/4$.

The Cheeger constant of a graph,
\begin{equation}
h=\min_{1\le|A|\le n/2}\frac{|\partial A|}{|A|},\qquad \frac{\lambda_2}2\le h\le\sqrt{2\Delta\lambda_2},
\label{eq:cheeger}
\end{equation}
with $\Delta$ the largest degree~\cite{mohar1989}, is out of reach at our sizes, so we restrict $A$ to arcs, the regions
of the contiguous distance [Eq.~(\ref{eq:dist}); Fig.~\ref{fig:graphs}(h)]:
\begin{equation}
h_{\mathrm{arc}}=\min_{\substack{A\text{ an arc}\\ 1\le|A|\le n/2}}\frac{|\partial A|}{|A|}\ \ge h.
\label{eq:harc}
\end{equation}
On the ring, $h_{\mathrm{arc}}=12/n$.

Our move is Olfati-Saber's random rewiring operator $R_p$~\cite[Sec.~IV]{olfatisaber2005} at range $k=2$ and
$p=\beta$, with the order of the visits and the end a bond keeps, which Ref.~\cite{olfatisaber2005} leaves open, fixed by
Eq.~(\ref{eq:order}).

\subsection{The rewiring angle}\label{sec:graphs-angle}

We parameterize the rewiring by $\theta=\arcsin\sqrt\beta$. The whole dependence on $\beta$ sits in the $2n$ attempt
bits, independent with probability $\beta$ of being 1, whose Fisher information~\cite{fisher1925,rao1945}
\begin{equation}
\mathrm{d}s^2=\frac{2n\,\mathrm{d}\beta^2}{\beta(1-\beta)}=8n\,\mathrm{d}\theta^2,\qquad 0\le\theta\le\pi/2,
\label{eq:fr}
\end{equation}
diverges at both ends in $\beta$ but is constant in $\theta$: equal steps in $\theta$ separate the laws of the bits
equally~\cite{amari2000,wootters1981}, and those of the graphs, a coarse-graining of the bits, at most as
far~\cite{cencov1982}. Our grid is uniform in $\theta$,
\begin{equation}
\theta_j=\frac{j\pi}{98},\qquad j=0,\ldots,49.
\label{eq:grid}
\end{equation}
Its first point past the ring, $\beta_1=\sin^2(\pi/98)=0.00103$, is a mean of $2n\beta_1=0.13$ rewiring attempts at
$n=64$ and 2.1 at $n=1024$.

\section{Circuits and simulations}\label{sec:sim}

Our circuits act on $n$ system qubits on the vertices of a graph $W(n,\beta)$ of Sec.~\ref{sec:graphs}. Each layer
applies $n/2$ random two-qubit Clifford gates on bonds of the graph and then measures each qubit with probability $p$;
a reference, entangled with the system at the start, records the logical qubits that survive.

Each system qubit $i$ starts in a Bell pair with a reference qubit $i'$,
\begin{equation}
|\psi(0)\rangle=\bigotimes_{i=0}^{n-1}\frac{|00\rangle_{ii'}+|11\rangle_{ii'}}{\sqrt2},
\label{eq:bell}
\end{equation}
so the system starts maximally mixed and purified by the reference~\cite{gullans2020prx,li2021}. Layer $t$ applies
$n/2$ two-qubit gates $u_{t,a}$ in sequence, each drawn uniformly from the $11{,}520$ two-qubit Clifford gates up to
phase and placed on a bond drawn uniformly, with replacement, from the $2n$ bonds of the graph $W(n,\beta)$, and then
measures each system qubit in the $Z$ basis independently with probability $p$:
\begin{equation}
|\psi(t)\rangle\propto M_t\,u_{t,n/2}\cdots u_{t,2}\,u_{t,1}\,|\psi(t-1)\rangle,\qquad
M_t=\prod_{i\in\mathcal M_t}\frac{1+s_{t,i}Z_i}{2},
\label{eq:layer}
\end{equation}
with $\mathcal M_t$ the measured qubits and $s_{t,i}=\pm1$ their outcomes. The gates follow the bonds of a fixed sparse graph,
nonlocal once the ring is rewired, and are drawn at random: within a layer a qubit $i$ takes
part in $\deg i/4$ gates on average, and possibly in several or none. A circuit of depth $T$ has $nT/2$ gates and on
average $pnT$ measurements. The gates and measurements are ideal, and the reference is never measured. Its entropy is the
number of logical qubits,
\begin{equation}
k(t)=S_R(t),
\label{eq:k}
\end{equation}
which never increases, and a region $A$ of the system holds the information
\begin{equation}
\IAR=S_A+S_R-S_{AR},\qquad 0\le\IAR\le2k,
\label{eq:iar}
\end{equation}
about the reference. The state stays a stabilizer state, so the circuits can be simulated
efficiently~\cite{aaronson2004} and every entropy follows from the stabilizer group~\cite{fattal2004}; none of these
entropies depends on the outcomes $s_{t,i}$.

Every simulation draws its graphs, gates, and measurements from the same master seeds (2026102017 for the graphs,
2026102029 for the gates, and 2026102039 for the measurements). Graph $g$ of a given $(n,\beta)$ is thus the same in
every simulation, its circuit has the same gates at every rate, and the qubits measured at a rate $p$ include those
measured at every lower rate; ``the same circuits'' below means the same graphs, gates, and measurement draws.
Table~\ref{tab:runs} lists the two main simulations. The rate $p\simeq p_c/2$ of the main text, $p=0.0728$ on the ring
and 0.1779 at $\beta=1$, is half of the preliminary estimates $p_c^{\mathrm{pin}}=0.1457$ and 0.3559 with which the
runs were set, and $0.503\,p_c$ and $0.473\,p_c$ of the final values (Sec.~\ref{sec:exp}); every rate written
$p=f\,p_c^{\mathrm{pin}}$ below uses these $p_c^{\mathrm{pin}}$.

\begin{table}[tbp]
\def\rb#1#2{\parbox[t]{#1}{\raggedright #2\strut}}\def\mk#1{\textsuperscript{\,#1}}\caption{The two simulations. Each condition $(n,\beta,p)$ runs one circuit on each of its graphs, for $T$ layers or
until it stops, and records the listed quantities at every layer; $\ell$ is the number of exposed qubits
(Sec.~\ref{sec:obs-codes}).}
\label{tab:runs}
\begin{ruledtabular}
\begin{tabular}{lllllllll}
Simulation & $n$ & $\beta$ & $p$ & Graphs & $T$, ring & $T$, $\beta=1$ & Stops at & Recorded \\ \hline
\rb{0.85in}{\textbf{Purification transition} (Fig.~\ref{fig:fss})} & $64,96,\ldots,256$ &
\rb{0.85in}{$\sin^2(j\pi/98)$, $j=0,\ldots,49$} & \rb{1.3in}{every 0.05, and every 0.005 near $p_c$\mk{a}} & 1000 &
\multicolumn{2}{l}{$16n$} & $k=0$ & $k$, $\dcont$ \\
 & 512, 1024 & & every 0.005 near $p_c$\mk{b} & 1000 & \multicolumn{2}{l}{$16n$} & & $k$ \\[3pt] \hline
\rb{0.85in}{\textbf{Coding front} [Fig.~\ref{fig:enc}(a), (b), and (d)]} & $64,96,\ldots,256$ & 0 and 1 &
\rb{1.3in}{$\simeq p_c/4$, $p_c/2$, and $3p_c/4$\mk{c}} & $10^5$ & 1024 & 256 & \rb{0.7in}{encoding or $k=0$} &
\rb{0.75in}{$k$, whether $\ell>0$} \\
 & 384--4096\mk{d} & & & $10^4$ & 1024 & 256 & & \\
 & 8192 & & & $10^3$ & 1536 & 512 & & \\
 & 16384 & & & 100 & 2048 & 512 & & \\
 & 32768 & & & 100 & 3072 & 512 & & \\
\end{tabular}
\end{ruledtabular}
\par\vspace{3pt}{\footnotesize\raggedright
\mk{a}Ring 0.105--0.25 at $n=64$ and 0.13--0.25 at $n=256$; $\beta=1$ 0.15--0.485 and 0.235--0.485. The ranges differ
by angle and size (Sec.~\ref{sec:fss-data}).\\
\mk{b}Ring 0.130--0.25 at $n=512$ and 0.140--0.195 at $n=1024$, with $p=0.1425$ and 0.1435 added at both sizes; $\beta=1$
0.275--0.485 and 0.295--0.45.\\
\mk{c}$p=0.0364$, 0.0728, and 0.1093 on the ring and 0.0890, 0.1779, and 0.2669 at
$\beta=1$: 0.25, 0.5, and 0.75 of $p_c^{\mathrm{pin}}$.\\
\mk{d}$n=384$, 512, 768, 1024, 1536, 2048, and 4096; at $\beta=1$ and $p\simeq p_c/2$, $10^5$ graphs at $n=512$, 1024,
2048, and 4096.\par}
\end{table}

\section{Observables}\label{sec:obs}

\subsection{Code properties}\label{sec:obs-codes}

A region $A\subseteq V$ is exposed when $\IAR>0$. The code distance $d$ is the size of the smallest exposed region, and
the contiguous distance $\dcont$ the length of the shortest exposed arc of consecutive qubits along the original
ring~\cite{li2021}:
\begin{equation}
d=\min\bigl\{|A|:\IAR>0\bigr\},\qquad \dcont=\min\bigl\{|A|:A\text{ an arc},\ \IAR>0\bigr\}\ge d.
\label{eq:dist}
\end{equation}
The number of exposed single qubits is
\begin{equation}
\ell=\bigl|\{i\in V:\ I(i{:}R)>0\}\bigr|,
\label{eq:ell}
\end{equation}
so $\ell>0$ exactly when $d=\dcont=1$. The decoupling size is the longest arc of the first qubits that is not exposed,
\begin{equation}
A^*=\max\bigl\{a:\ I(\{0,\ldots,a-1\}{:}R)=0\bigr\},
\label{eq:astar}
\end{equation}
and since $\IAR$ only grows with $A$, $\dcont\le A^*+1$. Earlier work reads the contiguous distance off the mean of $\IAR$ over
circuits~\cite{li2021,li2023dpre}; here $\dcont$, $\ell$, and $A^*$ are computed in each circuit and then averaged. These quantities are defined when $k>0$, and their averages are
taken over the circuits that still hold a code,
\begin{equation}
\avgc{X}=\frac{\sum_g X_g\,\mathbf 1[k_g>0]}{\sum_g\mathbf 1[k_g>0]},
\label{eq:cond}
\end{equation}
with $g$ running over the graphs of a condition.

\subsection{First passages and fronts}\label{sec:obs-times}

The first layers at which a circuit purifies, nearly purifies, and encodes are
\begin{equation}
\tau_0=\min\{t:k(t)=0\},\qquad \tau_1=\min\{t:k(t)\le1\},\qquad \tenc=\min\{t:k(t)>0,\ \ell(t)=0\}.
\label{eq:tau}
\end{equation}
A circuit that has not reached such a layer by the depth $T$ is censored, and medians are Kaplan--Meier
medians~\cite{kaplan1958}, undefined when more than half the circuits are censored. The encoding front, the encoding
time, the width of the front, and the scaled depth are
\begin{equation}
\Phi_n(T)=P(\tenc\le T),\qquad \Tenc=\med\,\tenc,\qquad w=\Phi_n^{-1}\bigl(\tfrac34\bigr)-\Phi_n^{-1}\bigl(\tfrac14\bigr),
\qquad u=\frac{T-\Tenc}{w},
\label{eq:front}
\end{equation}
and $T_{1/n}$ is the first layer at which $\avgc{\ell}<1$,
interpolated linearly between the two layers around the crossing. Unlike the Kaplan--Meier medians, $\Tenc$ is the
median of $\Phi_n$ over all circuits, interpolated linearly between layers: a circuit that purifies before it encodes
stays in the denominator rather than being censored.

The time $\tau_0$ is the purification time of the maximally mixed state~\cite{gullans2020prx}; $\tau_1$ is not the
purification time of a single reference qubit used there as an order parameter.

\subsection{Erasure}\label{sec:obs-erasure}

Erasing each qubit independently with probability $e$ loses logical information with probability
\begin{equation}
P_f(e)=\sum_{A\subseteq V}e^{|A|}(1-e)^{n-|A|}\,\mathbf 1[\IAR>0]=\sum_{m=0}^n\binom nm e^m(1-e)^{n-m}f_m,
\label{eq:pf}
\end{equation}
with $f_m$ the share of the $m$-qubit regions that are exposed: the failure probability of an optimal erasure
decoder~\cite{grassl1997,bennett1997}. At small $e$, $P_f\simeq N_d\,e^d$, with $N_d$ the number of exposed regions of
size $d$, and $e_{1/2}$ solves $P_f(e_{1/2})=1/2$. Section~\ref{sec:erasure-est} gives the estimator of $f_m$.

\subsection{Averages and errors}\label{sec:obs-errors}

Averages are over the graphs of a condition, one circuit each, and their errors are standard errors over graphs. Medians,
fronts, and fits carry 16--84\% intervals over 2000 resamples of whole graphs~\cite{efron1979} unless stated, with one
resample shared by all rates (and, in Sec.~\ref{sec:fss}, all angles) of a size. The errors of $p_c$, $\nu$, and $z$
(Sec.~\ref{sec:exp}) combine such resamples with systematic shifts, Eq.~(\ref{eq:err}), and the exponents $\alpha$ of
Sec.~\ref{sec:codes} and the width exponents of Sec.~\ref{sec:coding-front} carry the standard error of a weighted
least-squares fit.

\section{Phase diagrams}\label{sec:phase}

\begin{figure*}[tbp]
\centering
\includegraphics[width=\textwidth]{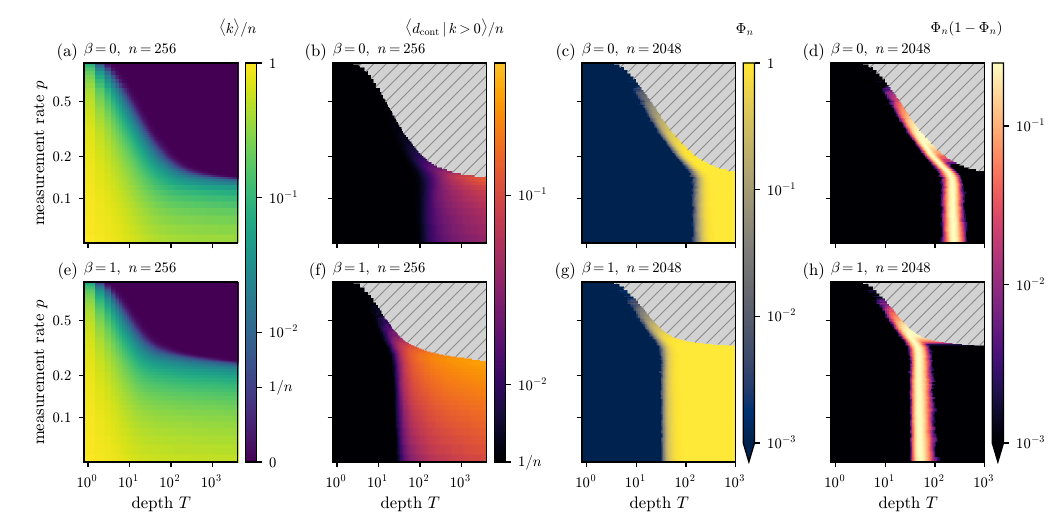}
\caption{The ring (top, $\beta=0$) and fully rewired graphs (bottom, $\beta=1$) on the plane of depth $T$ and measurement
rate $p$: (a),~(e)~the code rate $\langle k\rangle/n$ over all circuits (linear below $1/n$, logarithmic
above); (b),~(f)~the contiguous distance per qubit $\avgc{\dcont}/n$; (c),~(g)~the encoding front $\Phi_n=P(\tenc\le
T)$; (d),~(h)~its variance over circuits, $\Phi_n(1-\Phi_n)$; (a), (b), (e), and (f) at $n=256$, to $T=16n$; (c), (d),
(g), and (h) at $n=2048$, to $T=1000$; (b)--(d) and (f)--(h) on logarithmic scales, gray and hatched where
fewer than 10\% of the circuits hold a code, as in Figs.~\ref{fig:model}(c) and \ref{fig:enc}(c) of
the main text. Each column shares one color scale; values below $10^{-3}$ (none of the 1000 circuits) take the lowest
color. }
\label{fig:phase}
\end{figure*}

By $T=16n$, $\langle k\rangle/n$ has fallen below $1/n$ for $p\ge0.14$ on the ring, the rate of the grid next to
$p_c=0.1447$, but already for $p\ge0.25$ at $\beta=1$, well below $p_c=0.376$ (Sec.~\ref{sec:exp}): at finite $n$ the
reference also purifies below $p_c$, after a depth that grows with $n$ (Fig.~\ref{fig:phase}).
Figure~\ref{fig:model}(c) of the main text is the plane of fully rewired graphs at $n=256$ (95 rates from 0.01 to
0.95, 1000 graphs each, to $T=16n$), and Fig.~\ref{fig:phase} sets the ring beside it; each point $(T,p)$ holds one
circuit per graph, read at depth $T$. The mean $\avgc{\dcont}$ is over the circuits that hold a code, and the encoding
front $\Phi_n$ counts a circuit as encoded even after it is exposed again or purifies, so both are left out (gray and
hatched) where fewer than 10\% of the circuits hold a code. At $n=2048$ the encoding front rises from 0 to 1 at $\Tenc$
only at low rates: above $p=0.215$ on the ring and $p=0.36$ at $\beta=1$, part of the circuits purify before they
encode, and $\Phi_n$ levels off below 0.9.

\section{Codes at logarithmic depth}\label{sec:codes}

At logarithmic depth, fully rewired graphs give codes of finite rate with an extensive mean contiguous
distance, while on the ring the mean contiguous distance stays bounded. Figure~\ref{fig:codes}(a), (b) of the main
text reads at $T=8\log_2n$ (to the nearest layer: 48 at $n=64$, 64 at $n=256$) the circuits of the
purification-transition simulation (Table~\ref{tab:runs}) at $n=64$--256, one on each of 1000 graphs.
Figure~\ref{fig:codes}(a) adds 14 rates run on the same circuits (12 from $p=0.001$ to 0.045, and 0.075 and 0.125);
its main panel shows $n=256$ and its inset $n=96$--256. Its black circles and Fig.~\ref{fig:codes}(b) are at
$p\simeq p_c/2$ (Sec.~\ref{sec:sim}): $p=0.0728$ on the ring and 0.1779 at $\beta=1$, with $\IAR$ recorded for every
prefix $A=\{0,\ldots,a-1\}$, and, at $n=256$ only, $p=0.1225$, 0.1396, 0.1535, 0.1681, and 0.1705 at $\theta_{10}$,
$\theta_{13}$, $\theta_{18}$, $\theta_{23}$, and $\theta_{30}$. Every average is over the circuits that hold a code,
Eq.~(\ref{eq:cond}).

\subsection{Rate and contiguous distance}\label{sec:codes-rate}

The depth $T=8\log_2n$ lies past the encoding time at $\beta=1$ but before it on the ring: at $p\simeq p_c/2$ and
$n=64$--256, $\Tenc=28.0$--35.8 at $\beta=1$ against $T=48$--64, but $\Tenc=54.3$--96.3 on the ring
(Sec.~\ref{sec:coding-n}). Most of the ring's codes therefore still have an exposed qubit, and their contiguous distance
is one qubit (Sec.~\ref{sec:codes-d}).

\begin{figure}[tbp]
\centering
\includegraphics[width=\textwidth]{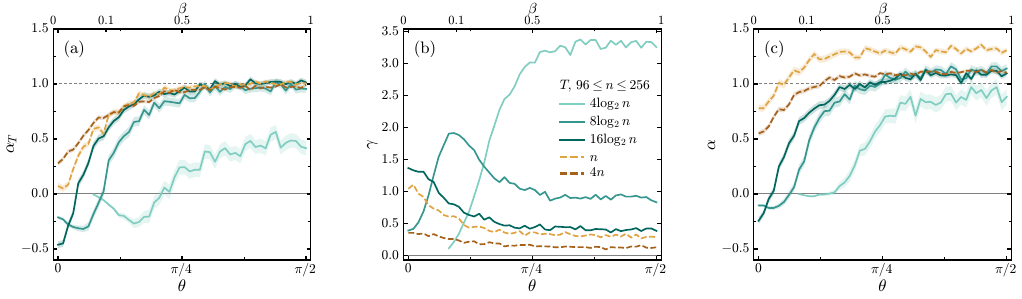}
\caption{The exponent of the mean contiguous distance against the angle $\theta$ ($\beta$ on the top axis), at the code
rate $r=0.25$ and over $n=96$--256, at three logarithmic depths, $T=4$, 8, and $16\log_2n$ (teal, solid), and two linear
ones, $T=n$ and $4n$ (amber, dashed). (a)~The exponent $\alpha_T$ at a depth shared by all sizes, each family's depth at
$n=256$: $T=32$, 64, 128, 256, and 1024. (b)~The growth of the codes with depth, $\gamma$, along each family, averaged over
the sizes. (c)~The exponent $\alpha$ along each family, Eq.~(\ref{eq:dexp}). Bands: the error of the weighted fit. At
$T=4\log_2n$ the smallest angles have no fit, since nearly every code has $\dcont=1$. Gray dashed lines: an extensive mean
contiguous distance.}
\label{fig:alpha}
\end{figure}

Where a curve crosses the code rates $r=0.25$ and 0.1 we read $\avgc{\dcont}/n$, interpolating linearly between
neighboring rates, and fit
\begin{equation}
\avgc{\dcont}\propto n^{\alpha}
\label{eq:dexp}
\end{equation}
over $n=96$--256, the sizes of the inset of Fig.~\ref{fig:codes}(a): $\alpha=1$ is an extensive mean contiguous
distance, $\alpha=0$ a bounded one. At $T=8\log_2n$, $\alpha=1.14(4)$ at $\beta=1$ (both rates) and $-0.10(1)$ and
$-0.17(1)$ on the ring (at $r=0.25$ and 0.1; the figures show $r=0.25$), and it grows with $\beta$ in between
[Fig.~\ref{fig:alpha}(c)].

\begin{figure}[tbp]
\centering
\includegraphics[width=\textwidth]{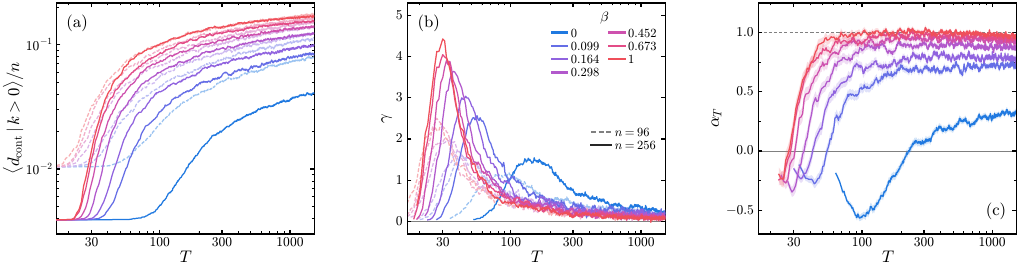}
\caption{The mean contiguous distance against the depth $T$ at the code rate $r=0.25$, at the seven angles of
Fig.~\ref{fig:codes} of the main text (colors as there). (a)~$\avgc{\dcont}/n$ and (b)~its growth with depth, $\gamma$, at $n=96$
(light, dashed) and 256 (solid). (c)~The exponent $\alpha_T$ at a depth shared by all sizes, fitted over $n=96$--256;
bands: the error of the weighted fit. Gray dashed line: an extensive mean contiguous distance.}
\label{fig:alphaT}
\end{figure}

At a fixed depth $T$ shared by all sizes, the exponent $\alpha_T$ at $\beta=1$ stays between 0.91 and 1.04 from $T=48$
to 1536, while on the ring it is at most 0.35 [Figs.~\ref{fig:alpha}(a) and \ref{fig:alphaT}(c)]. When the depth grows
with $n$, $\alpha$ also counts the growth of each code with depth,
$\gamma=\mathrm{d}\ln\avgc{\dcont}/\mathrm{d}\ln T$ [Figs.~\ref{fig:alpha}(b) and \ref{fig:alphaT}(a), (b)]:
\begin{equation}
\alpha\simeq\alpha_T+\gamma\,\frac{\mathrm{d}\ln T}{\mathrm{d}\ln n}.
\label{eq:alphaT}
\end{equation}
The factor $\mathrm{d}\ln T/\mathrm{d}\ln n$ is 1 for $T\propto n$ but about $1/\ln n\approx0.2$ for $T\propto\log n$. At
$T=n$ the codes still grow with depth ($\gamma=0.29$ at $\beta=1$), which lifts $\alpha$ above 1.

\subsection{Individual codes}\label{sec:codes-d}

Individual codes differ widely from the mean. We take the 1000 codes of the erasure simulation at each of its seven
rates per end ($n=256$, $T=8\log_2n=64$; Sec.~\ref{sec:erasure}), at $p\simeq p_c/2$ the same circuits as above, with $d$
found among all regions of up to four qubits (Sec.~\ref{sec:erasure-est}). On the ring no code has $d>2$ at any rate, and at most 15 of the 1000 have $d=2$. At $\beta=1$ and
$p\simeq p_c/2$, 348 codes still have $d=1$ and 78.2\% have $d\le4$, while $\langle\dcont\rangle=25.6$ (median 26). The
contiguous distance overstates $d$ by an order of magnitude: of the 175 codes with $d=2$, only 6 have an exposed arc of
two qubits; in the others the exposed pairs are not neighbors on the ring.

\subsection{Decoupling along an arc}\label{sec:codes-astar}

Figure~\ref{fig:codes}(b) grows the arc $A=\{0,\ldots,a-1\}$ and averages $\IAR/2k$, circuit by circuit, over the
circuits that hold a code. At $\beta=1$ the curves nearly collapse: at $n\ge160$ they lie within 0.0082 of the curve at
$n=256$ over all $|A|/n$ (0.0131 on the ring). Short arcs hold almost no information, and less as $n$ grows: an arc of
$0.1n$ qubits exposes a share 0.0124 at $n=64$ and 0.0032 at $n=256$. On the ring the same share grows with $n$, from
0.0499 to 0.0926: short arcs already expose information.

\begin{figure*}[tbp]
\centering
\includegraphics[width=\textwidth]{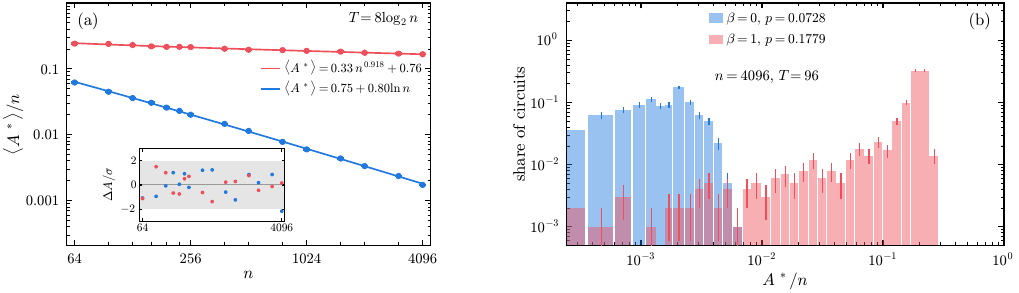}
\caption{The decoupling size $A^*$ at $p\simeq p_c/2$ ($p=0.0728$ on the ring, blue; 0.1779 at $\beta=1$, red).
(a)~$\langle A^*\rangle/n$ against $n$ at $T=8\log_2n$; lines: $\langle A^*\rangle=c\,n^a+d$ at $\beta=1$ and
$\langle A^*\rangle=a+b\ln n$ on the ring; inset: their residuals in units of the standard error, shaded within $\pm2$.
(b)~The distribution of
$A^*/n$ over the 1000 circuits at $n=4096$ and $T=96$, on logarithmic bins of whole qubits; the circuits with
$A^*=0$ (21 on the ring, 2 at $\beta=1$) are not shown. Error bars: one standard error over
graphs, mostly smaller than the points.}
\label{fig:astar}
\end{figure*}

The decoupling size $A^*$ [Eq.~(\ref{eq:astar})] is the longest such arc that holds no information, and
$\dcont\le A^*+1$ in every circuit. Figure~\ref{fig:astar} extends it to $n=4096$, with the same ensemble, rates, and
depths at $n=384$--4096, 1000 graphs each. At $\beta=1$, $\langle A^*\rangle$ grows nearly, but not exactly, in
proportion to $n$: $\langle A^*\rangle/n$ falls from 0.242 at $n=64$ to 0.167 at $n=4096$ at $T=8\log_2n$
[Fig.~\ref{fig:astar}(a)], and from 0.362 to 0.204 at $T=32\log_2n$. The form $\langle A^*\rangle=c\,n^a+d$ fits both
depths, with $a=0.918$ and 0.905 ($\chi^2/\mathrm{dof}=0.80$ and 0.78); fitted to $n\le2048$, it predicted $n=3072$ and
4096 within 1.1 standard errors, whereas an extensive decoupling size, a constant $\langle A^*\rangle/n$ with a
power-law correction, missed $n=4096$ at $T=32\log_2n$ by 3.3. At $T=8\log_2n$, the depth of the main text, the
extensive form fits as well ($\langle A^*\rangle/n\to0.10$, $\chi^2/\mathrm{dof}=0.76$): there the data do not separate
the two. On the ring, $\langle A^*\rangle$ grows only as $\ln n$: $\langle A^*\rangle=0.75+0.80\ln n$ at $T=8\log_2n$
($\chi^2/\mathrm{dof}=1.16$).

Individual circuits spread widely [Fig.~\ref{fig:astar}(b)]: at $n=4096$ and $\beta=1$, $A^*/n$ has mean 0.167, median
0.189, and standard deviation 0.064.

\subsection{Domain-wall fluctuations and phenomenological Gaussian height model for the effective tension.}
\label{sec:codes-gauss}

Here we discuss a phenomenological model for domain-wall fluctuations in small-world random circuits. This model supports our conjecture in the main text that an extensive decoupling subregion $A^*$ can be explained by an extensive loss of
entropy upon pinning.

To understand the extensive crossover between pinned and
depinned domain walls, we consider an effective ferromagnetic
Ising model on the spacetime graph $G\times[0,T]$,
\begin{equation}
H_{\mathrm{Ising}}
=-J_t\sum_{i,t}\sigma_{i,t}\sigma_{i,t+1}
-J_s\sum_{i < j} A_{ij} \sigma_{i,t}\sigma_{j,t}.
\end{equation}
where $\sigma_{i,t} = \pm 1$ and $A_{ij}$ is the adjacency matrix of the small-world network $W(n,\beta)$.
Such statistical-mechanics descriptions arise naturally from
replica mappings of random unitary and monitored quantum
circuits~\cite{zhou2019emergent,jian2020,nahum2021}.
For the second R\'enyi entropy $S_{AR}^{(2)}$, the
reference system $R$ imposes uniform spin boundary conditions
at the bottom of spacetime, while the top boundary has
opposite orientations on $A$ and $A_c$.
We choose $s=+1$ on $R$ and $A$, and $s=-1$ on $A_c$.
These boundary conditions admit two competing interface
configurations: a pinned wall surrounding $A_c$, or a
wall surrounding $A$ together with a freely fluctuating
bulk wall separating the top and bottom boundaries.

In the ordered phase, we model a system-spanning domain wall
by a height field $h_i$ specifying the time coordinate at
which the interface crosses vertex $i$.
For a single-valued interface, the excess energy is
\begin{equation}
E[h]=\mu n+
J\sum_{i < j} A_{ij} |h_i-h_j|,
\label{eq:height_energy}
\end{equation}
where $\mu$ is the energy per vertex of a flat interface,
and $J$ penalizes height differences between connected
vertices. This representation follows by counting the
frustrated spatial bonds between neighboring worldlines.
At large circuit depth, we approximate the discrete height
coordinate by a continuous variable, appropriate for
describing fluctuations on scales much larger than one
circuit layer.

At long wavelengths, we approximate the interface by a
Gaussian elastic model,
\begin{equation}
E_{\mathrm G}[h]
=\mu n+\frac{\kappa}{2}
\sum_{i < j} A_{ij} (h_i-h_j)^2
=\mu n+\frac{\kappa}{2}\mathbf h^T L \mathbf h,
\label{eq:gaussian_height}
\end{equation}
where $L$ is the graph Laplacian, $\kappa$ is the
effective interfacial stiffness, and $\mathbf h = (h_1, h_2, \ldots, h_n)$ are the height coordinates.

Let $0=\lambda_1<\lambda_2\leq\cdots\leq\lambda_n$
denote the eigenvalues of $L$. Expanding the height
field in Laplacian eigenmodes reduces the partition
function to independent Gaussian integrals.
The zero mode corresponds to rigid translation of the
interface along the time direction, contributing a factor
proportional to $T$. The remaining $n-1$ modes give
\begin{equation}
Z_{\mathrm{bulk}}
=
e^{-\mu n}\,
\frac{T}{\tau}\sqrt{n}
\left(\frac{2\pi}{\kappa \tau^2}\right)^{(n-1)/2}
\prod_{\alpha=2}^{n}\lambda_\alpha^{-1/2},
\label{eq:bulk_partition}
\end{equation}
where $\tau$ specifies the normalization of the
continuum height measure. This expression assumes that
$T$ is sufficiently large that confinement of the
interface can be neglected. The corresponding bulk
free energy contains the spectral contribution
\begin{equation}
F_{\mathrm{bulk}}
=\frac{1}{2}\log\det' L_G
=\frac{1}{2}\sum_{\alpha=2}^{N}\log\lambda_\alpha,
\label{eq:bulk_spectral_free_energy}
\end{equation}
in addition to the flat-wall energy, translational
entropy, and Gaussian normalization terms.

To describe fluctuations of domain walls constrained by boundary pinning,
we fix $h_i=0$ for $i\in A$. The remaining heights
$h_i$, $i\in A_c$, are governed by the \emph{grounded Laplacian}
\begin{equation}
L_{A_c}=(L)_{A_c,A_c},
\label{eq:grounded_laplacian}
\end{equation}
obtained by deleting the rows and columns associated
with $A$. Crucially, this operation retains the energetic
contributions from edges connecting $A$ and $A_c$.
The Gaussian fluctuation partition function in this case is therefore
\begin{equation}
Z_{\mathrm{pin}}
=
e^{-\mu|A_c|}
\left(\frac{2\pi}{\kappa \tau^2}\right)^{|A_c|/2}
\left(\det L_{A_c}\right)^{-1/2}.
\label{eq:pinned_partition}
\end{equation}
Unlike the bulk interface, the grounded system has no
translational zero mode. Its spectral free-energy
contribution is
\begin{equation}
F_{\mathrm{pin}}
=\frac{1}{2}\log\det L_{A_c}
=\frac{1}{2}\sum_{\alpha=1}^{|A_c|}
\log\widetilde{\lambda}_\alpha,
\label{eq:pinned_spectral_free_energy}
\end{equation}
where $\widetilde{\lambda}_\alpha$ denote the strictly
positive eigenvalues of the grounded Laplacian.

We quantify the difference in the spectral contributions
to the effective interface tensions by defining
\begin{equation}
\mathcal D(\beta,a,N)
=
\frac{1}{(1-a)N}
\sum_{\alpha=1}^{(1-a)N}
\log\widetilde{\lambda}_\alpha
-
\frac{1}{N}
\sum_{\alpha=2}^{N}\log\lambda_\alpha,
\label{eq:dwall}
\end{equation}
where $a=|A|/N$ and $\beta$ denotes the small-world
rewiring probability. In the Gaussian approximation,
$\mathcal D/2$ gives the difference between the spectral
contributions to the pinned and bulk interface tensions.
The full physical tension difference additionally depends
on microscopic interface energetics and the normalization
of the continuum theory.

We evaluate $\mathcal D(\beta,a,N)$ numerically for
different system sizes, as shown in Fig.~2(c) of the main text.
For the unrewired ring ($\beta=0$), pinning affects
only a subextensive number of spatial bonds, and we
find $\mathcal D\to0$ as $N\to\infty$.
By contrast, in the fully rewired network ($\beta=1$),
an extensive number of edges connect $A$ and $A_c$.
Pinning therefore constrains fluctuations throughout
the interface, producing a finite difference in the
spectral free-energy densities:
$\mathcal D\to\mathrm{const}$ as $N\to\infty$.

This provides a mechanism by which long-range connectivity
can generate distinct effective tensions for pinned and
bulk domain walls, thereby enabling a crossover
at an extensive subsystem size $A^* \propto n$.

We evaluate Eq.~(\ref{eq:dwall}) exactly on every graph. Both determinants are of sparse positive-definite matrices,
$\det'\mathcal L$ as $n$ times the determinant of $\mathcal L$ without its row and column 0, and a sparse LU
factorization gives their logarithms as sums over its pivots, each checked positive. On graphs of $n\le256$ these agree
with dense determinants within $3.4\times10^{-13}$, and $\mathcal D$ within $6.6\times10^{-15}$; on the ring, whose
number of spanning trees is $nF_n^2$ with $F_n$ the $n$th Fibonacci number, $\ln\det'\mathcal L=\ln(n^2F_n^2)$ holds
within $1.1\times10^{-10}$ up to $n=32768$. The graphs are those of the simulations, drawn from the master graph seed
(Sec.~\ref{sec:data}): at $\beta=1$, every graph the simulations drew, $10^3$ to $10^5$ per size; at the five
intermediate angles of Fig.~\ref{fig:codes}, 1000 per size, with new graphs drawn the same way at $n=768$ and
1536--32768, where no circuit ran; and on the ring, where every graph is the same, one. Each of the 1,218,017 graphs
was checked connected, with $2n$ distinct bonds and no self-loops. Error bars are one standard error over graphs.

On the ring $\mathcal D<0$ and $\mathcal D\to0$, from $-0.0428$ at $n=64$ to $-2.5\times10^{-4}$ at $n=32768$: pinning
constrains only the bonds at the two ends of $A$, a number that does not grow with $|A|$, and costs no entropy per height
in the limit. Every rewired angle crosses zero, at a size that falls as $\beta$ grows (between $n=512$ and 768 at
$\beta=0.099$, between 64 and 96 at $\beta=1$), and then levels off at a positive value that grows with $\beta$: at
$n=32768$, $\mathcal D=0.00395(1)$ at $\beta=0.099$ and 0.02165(2) at $\beta=1$. The insets of
Fig.~\ref{fig:codes}(c) draw the two ends schematically, with 16 qubits and $|A|=4$: on the ring the 6 bonds between
$A$ and the rest meet 4 of the other 12 qubits; at $\beta=1$ the 15 such bonds meet 9.

\section{Erasure}\label{sec:erasure}

Codes from fully rewired graphs fail far less often under random erasure than those of the ring, at the same
$p/p_c$ and at the same code rate. Figure~\ref{fig:codes}(d) of the main text and this section use $n=256$ and
$T=8\log_2n=64$, with one circuit on each of 1000 graphs, at seven rates per end, $p=f\,p_c^{\mathrm{pin}}$ with
$f=0.1,0.2,\ldots,0.7$ (Sec.~\ref{sec:sim}; Table~\ref{tab:erasure}); $f=0.5$ is the rate $p\simeq p_c/2$ of the main
text. Every circuit holds a code at $T=64$.

\subsection{The estimator}\label{sec:erasure-est}

The failure probability of Eq.~(\ref{eq:pf}) needs the share $f_m$ of exposed $m$-qubit regions at every $m$. For
$m\le4$ we test every region, $177{,}589{,}056$ per circuit, so $f_1,\ldots,f_4$ and the distance $d\le4$ are exact.
Above, we draw $N_\pi=10^6$ uniformly random orderings $\pi$ of the qubits per circuit and record the size
$D(\pi)=\min\{m:\ I(\{\pi_1,\ldots,\pi_m\}{:}R)>0\}$ of the first exposed prefix.  The first $m$ qubits of a random
ordering are a uniformly random $m$-qubit region, and a region that contains an exposed one is exposed, so
\begin{equation}
f_m=P_\pi\bigl[D(\pi)\le m\bigr],\qquad \hat f_m=\frac1{N_\pi}\sum_{l=1}^{N_\pi}\mathbf 1\bigl[D(\pi_l)\le m\bigr]\quad(m>4),
\label{eq:orderings}
\end{equation}
and $\hat P_f(e)$ follows from Eq.~(\ref{eq:pf}) with the exact $f_m$ below and $\hat f_m$ above $m=4$. Failures too
rare to appear among the $N_\pi$ orderings are bounded by the rule of three: they add, at 95\% confidence,
at most $U(e)=(3/N_\pi)\,P[\mathrm{Bin}(n,e)\ge5]$, $3.5\times10^{-7}$ at $e=10^{-2}$. As a check, $10^5$ literal erasure
patterns per circuit, at ten values of $e$ from 0.01 to 0.4, agree with $\hat P_f(e)$ within $|z|\le3.2$ on the ring and
$1.5$ at $\beta=1$ (the largest over the ten values). Means over circuits carry standard errors over graphs, and ratios
their propagated errors.

\subsection{Results}\label{sec:erasure-res}

At $p\simeq p_c/2$ and $e=10^{-2}$ the mean failure probability is $0.0956(18)$ on the ring and $0.0140(9)$ at
$\beta=1$, a ratio of $6.82(47)$. Both means are linear in $e$ at small $e$, because the codes with an exposed qubit
fail at first order: those with $d=1$ carry 99.97\% of the ring's mean at $e=10^{-2}$ and 99.54\% of that at $\beta=1$.
Restricted to the codes with $d\ge2$, 15 on the ring and 652 at $\beta=1$, the means are $1.77(26)\times10^{-3}$ and
$9.9(11)\times10^{-5}$, a ratio of $17.8(32)$. Near $e=10^{-4}$, the local exponent $\mathrm{d}\ln P_f/\mathrm{d}\ln e$
of the median code is 1.0005 on the ring and 2.0001 at $\beta=1$.

The two ends reach different code rates at the same $p/p_c$ (Table~\ref{tab:erasure}), so we also compare them at
equal code rate: the mean $e_{1/2}$ at $\beta=1$, interpolated linearly in the code rate between its seven rates, is
2.43--2.54 times the ring's at each of the six ring rates inside that range, code rates 0.176--0.568.

\begin{table}[tbp]
\caption{Erasure at $n=256$ and $T=64$: the rates $p=f\,p_c^{\mathrm{pin}}$, also in units of the final $p_c$, the mean
code rate, the share of codes with $d=1$, the mean failure probability at $e=10^{-2}$, and the mean $e_{1/2}$, with
standard errors over the 1000 graphs.}
\label{tab:erasure}
\begin{ruledtabular}
\begin{tabular}{llllllll}
 & $f$ & $p$ & $p/p_c$ & $\langle k\rangle/n$ & $d=1$ & $\langle P_f(10^{-2})\rangle$ & $\langle e_{1/2}\rangle$ \\ \hline
$\beta=0$ & 0.1 & 0.0146 & 0.101 & 0.709 & 1.000 & 0.3094(40) & 0.01957(27) \\
 & 0.2 & 0.0291 & 0.201 & 0.568 & 0.998 & 0.1895(31) & 0.03280(48) \\
 & 0.3 & 0.0437 & 0.302 & 0.461 & 0.999 & 0.1431(25) & 0.04290(60) \\
 & 0.4 & 0.0583 & 0.403 & 0.373 & 0.991 & 0.1164(22) & 0.05213(73) \\
 & 0.5 & 0.0728 & 0.503 & 0.297 & 0.985 & 0.0956(18) & 0.06183(83) \\
 & 0.6 & 0.0874 & 0.604 & 0.232 & 0.991 & 0.0838(16) & 0.06931(88) \\
 & 0.7 & 0.1020 & 0.705 & 0.176 & 0.989 & 0.0742(14) & 0.07708(95) \\
\hline
$\beta=1$ & 0.1 & 0.0356 & 0.095 & 0.649 & 0.513 & 0.0607(35) & 0.06473(90) \\
 & 0.2 & 0.0712 & 0.189 & 0.457 & 0.437 & 0.0319(20) & 0.1057(13) \\
 & 0.3 & 0.1068 & 0.284 & 0.317 & 0.384 & 0.0232(17) & 0.1432(17) \\
 & 0.4 & 0.1424 & 0.379 & 0.211 & 0.362 & 0.0180(13) & 0.1803(21) \\
 & 0.5 & 0.1779 & 0.473 & 0.132 & 0.348 & 0.01401(94) & 0.2151(24) \\
 & 0.6 & 0.2135 & 0.568 & 0.074 & 0.322 & 0.00924(62) & 0.2582(26) \\
 & 0.7 & 0.2491 & 0.663 & 0.034 & 0.315 & 0.00738(47) & 0.2995(27) \\
\end{tabular}
\end{ruledtabular}
\end{table}

\section{The coding transition}\label{sec:coding}

On the ring the encoding time grows as a power of $n$; on fully rewired graphs it grows as $\ln n$.
Figure~\ref{fig:enc}(a), (b), and (d) of the main text and Secs.~\ref{sec:coding-n} and \ref{sec:coding-front} use
the coding-front simulation (Table~\ref{tab:runs}): 17 sizes from $n=64$ to 32768, at $p=f\,p_c^{\mathrm{pin}}$ with
$f=0.25$, 0.5, and 0.75 (Sec.~\ref{sec:sim}). Each circuit runs until it encodes or purifies, or to a horizon well past
$\Tenc$; none reached the horizon unencoded. Figure~\ref{fig:enc}(c) and Sec.~\ref{sec:coding-exp} use $n=2048$ at
$\beta=1$, 1000 graphs, and $p=0.005$--0.95 in steps of 0.005, every layer to $T=1000$; Sec.~\ref{sec:coding-beta} uses
the nine sizes $n=64,96,\ldots,256$, 384, and 512 at all 50 angles, 1000 graphs, and $p$ in steps of 0.005, to $T=16n$.

\subsection{Encoding time against size}\label{sec:coding-n}

\begin{figure*}[tbp]
\centering
\includegraphics[width=\textwidth]{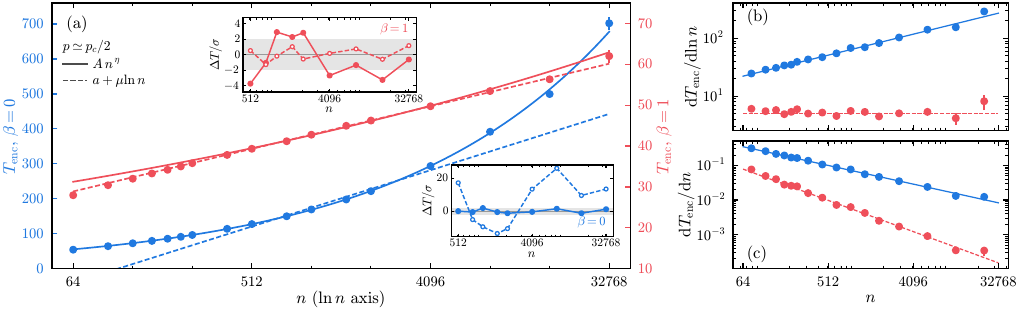}
\caption{The encoding time against size at $p\simeq p_c/2$ ($p=0.0728$ on the ring, blue, left axis; 0.1779 at
$\beta=1$, red, right axis). (a)~$\Tenc$ against $\ln n$ on linear axes, on which a logarithm is a straight line; lines:
the fits of Eq.~(\ref{eq:tencfits}) over $n\ge512$, drawn from $n=64$, the power law solid and
the logarithm dashed. Insets: the residuals $\Delta T$ of both fits over the fitted sizes, in units of $\sigma$, half the
16--84\% range; shaded, $|\Delta T|\le2\sigma$. (b)~The local slope $\Delta\Tenc/\Delta\ln n$ and (c)~$\Delta\Tenc/\Delta
n$ between neighboring sizes, at their geometric mean; lines: the derivatives of the fits, $\eta A\,n^{\eta}$ and $\eta
A\,n^{\eta-1}$ on the ring, $\mu$ and $\mu/n$ at $\beta=1$. Error bars: 16--84\% over 1000 graph resamples
(Table~\ref{tab:tencn}), propagated to the differences in (b) and (c).}
\label{fig:tenc}
\end{figure*}

We fit the encoding times over the nine sizes $n\ge512$, weighted by their variances over the graph resamples, with a
power law and a logarithm,
\begin{equation}
\Tenc=A\,n^{\eta},\qquad \Tenc=a+\mu\ln n,
\label{eq:tencfits}
\end{equation}
and, as checks weighted by the 16--84\% intervals, with $\Tenc=a+c\sqrt n$, $\Tenc=c\,(\ln n)^{\kappa}$, and
$\Tenc=a+c\,n^{\eta}$ with $\eta$ free. At $p\simeq p_c/2$ the two ends select different forms
(Fig.~\ref{fig:tenc}). On the ring the power law fits, $\eta=0.4018$ [0.3986, 0.4049] (95\%) with
$\chi^2/\mathrm{dof}=1.57$, against 251 for the logarithm, 17.3 for $\sqrt n$, and 45.3 for a power of $\ln n$. At
$\beta=1$ the logarithm fits, $\mu=5.01$ [4.96, 5.06] with $\chi^2/\mathrm{dof}=0.81$, against 7.28 for the power law and
194 for $\sqrt n$; $(\ln n)^\kappa$ with $\kappa=0.82$ also fits ($\chi^2/\mathrm{dof}=1.63$), and with $\eta$ free the
best $\eta$ lies at the edge of the scan, $\eta\to0$, where $a+c\,n^\eta$ becomes a logarithm. The local exponent
$\eta_{\mathrm{eff}}=\ln[\Tenc(n_2)/\Tenc(n_1)]/\ln(n_2/n_1)$ between neighboring sizes agrees: on the ring it stays at
0.38--0.43 up to $n=8192$, with no drift toward $1/2$, while at $\beta=1$ it falls from 0.21 to 0.098 as the local slope
$\mu_{\mathrm{eff}}=\Delta\Tenc/\Delta\ln n$ stays at 4.5--6.0. At $n=32768$, $\Tenc=702$ on the ring and 62 at
$\beta=1$; Table~\ref{tab:tencn} lists $\Tenc$ at $p\simeq p_c/2$ at every size.

The ring's exponent rises with the rate, from 0.391 at $f=0.25$ to 0.421 at $f=0.75$. At $\beta=1$ the logarithm fits at
all three rates; at $f=0.25$, where $\Tenc$ grows least, a power law with $\eta=0.090$ fits as well. Gullans et al.\ find
that random local circuits approach the erasure capacity at a fixed gap at a depth $O(\log n)$ in any dimension, but
need $O(\sqrt n)$ in one dimension when the gap closes with $n$~\cite{gullans2021}; their circuits are unitary encoders
judged by the failure probability, so these depths are not our $\Tenc$.

\begin{table}[tbp]
\caption{The encoding time at $p\simeq p_c/2$ ($p=0.0728$ on the ring, 0.1779 at $\beta=1$) at every size,
with the graphs of each condition and the 16--84\% range over 1000 graph resamples. At $\beta=1$ and $n=512$, 1024,
2048, and 4096 the $10^5$ graphs are those of Fig.~\ref{fig:enc}(a), (b).}
\label{tab:tencn}
\begin{ruledtabular}
\begin{tabular}{lllll}
 & \multicolumn{2}{c}{ring, $\beta=0$} & \multicolumn{2}{c}{$\beta=1$} \\
$n$ & graphs & $\Tenc$ & graphs & $\Tenc$ \\ \hline
64 & $10^5$ & 54.29 [54.22, 54.35] & $10^5$ & 27.97 [27.94, 28.01] \\
96 & $10^5$ & 64.29 [64.22, 64.37] & $10^5$ & 30.42 [30.39, 30.46] \\
128 & $10^5$ & 72.57 [72.49, 72.65] & $10^5$ & 32.01 [31.97, 32.05] \\
160 & $10^5$ & 79.46 [79.39, 79.55] & $10^5$ & 33.29 [33.26, 33.33] \\
192 & $10^5$ & 85.68 [85.60, 85.78] & $10^5$ & 34.19 [34.15, 34.22] \\
224 & $10^5$ & 91.09 [90.99, 91.19] & $10^5$ & 35.01 [34.98, 35.05] \\
256 & $10^5$ & 96.33 [96.23, 96.42] & $10^5$ & 35.81 [35.77, 35.84] \\
384 & $10^4$ & 113.88 [113.50, 114.24] & $10^4$ & 37.83 [37.72, 37.95] \\
512 & $10^4$ & 127.47 [127.09, 127.89] & $10^5$ & 39.33 [39.29, 39.37] \\
768 & $10^4$ & 149.78 [149.36, 150.24] & $10^4$ & 41.18 [41.04, 41.30] \\
1024 & $10^4$ & 169.39 [168.88, 169.91] & $10^5$ & 42.78 [42.74, 42.81] \\
1536 & $10^4$ & 197.89 [197.33, 198.49] & $10^4$ & 44.95 [44.82, 45.08] \\
2048 & $10^4$ & 221.75 [221.19, 222.36] & $10^5$ & 46.23 [46.20, 46.28] \\
4096 & $10^4$ & 293.59 [292.64, 294.42] & $10^5$ & 49.74 [49.70, 49.78] \\
8192 & $10^3$ & 391.64 [389.69, 394.00] & $10^3$ & 53.45 [53.11, 53.76] \\
16384 & 100 & 499.5 [484.0, 507.0] & 100 & 56.33 [55.67, 56.85] \\
32768 & 100 & 702.0 [681.7, 720.0] & 100 & 62.0 [60.0, 63.1] \\
\end{tabular}
\end{ruledtabular}
\end{table}

\subsection{The fronts}\label{sec:coding-front}

The front sharpens relative to $\Tenc$, but slowly. Its width $w=T_{3/4}-T_{1/4}$ [Eq.~(\ref{eq:front}), with
$\Phi_n(T_q)=q$] grows with $n$, but more slowly than $\Tenc$: over $n\ge512$, $w/\Tenc\propto n^{-0.086(6)}$ on the
ring and $n^{-0.075(2)}$ at $\beta=1$ ($\chi^2/\mathrm{dof}=0.84$ and 0.64).

\subsection{Exposure after encoding}\label{sec:coding-exp}

At $n=2048$ and $\beta=1$ the encoding time stays between 45.7 and 49.8 layers for $p=0.05$--0.30, and rises to 58.0
at $p=0.01$; the mean $\avgc{\ell}$ at $2\Tenc$ is 8.29 at $p=0.05$, 2.72 at 0.15, and 1.33 at 0.25.

Past $T_{1/n}$ (Sec.~\ref{sec:obs-times}), $\avgc{\ell}$ can rise above 1 again, and both the first dip and the last
crossing come later as $p$ falls. From $p=0.14$ to 0.33, $\avgc{\ell}$ stays below 1 within 35\% of $T_{1/n}$ (from
167.8, and for good after 203.8, at $p=0.15$); at $p=0.05$ it first dips below 1 at $T_{1/n}=410.5$ and stays below only
after layer 957.1. The 16--84\% interval of $T_{1/n}$ over 200 graph resamples, 283.7--334.1 at $p=0.05$, lies below the
dip, since resampled means dip earlier.

\subsection{Between the ends}\label{sec:coding-beta}

\begin{figure}[tbp]
\centering
\includegraphics[width=\textwidth]{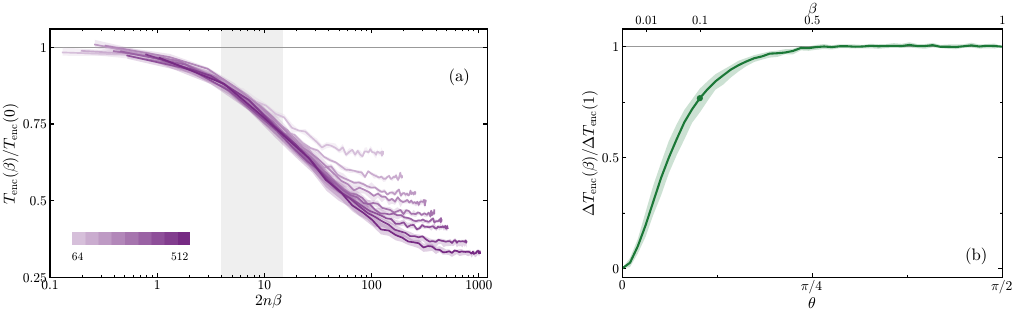}
\caption{The encoding time between the ends, over $n=64$--512 and $p=0.01$--0.10. (a)~$\Tenc(\beta)/\Tenc(0)$
against the number of rewired bonds $2n\beta$ for each size (light to dark, key): median (line) and quartiles (band)
over the 19 rates. Gray: the numbers of rewired bonds at which $\lambda_2$ doubles and the mean path length halves on
the graphs of Fig.~\ref{fig:graphs}. (b)~The share of the drop made by $\beta$, $\Delta\Tenc(\beta)/\Delta\Tenc(1)$
with $\Delta\Tenc(\beta)=\Tenc(0)-\Tenc(\beta)$, against the angle $\theta$ ($\beta$ on top): median (line) and
quartiles (band) over the 171 pairs of size and rate; dot: $\beta=0.099$.}
\label{fig:tencbeta}
\end{figure}

A few rewired bonds already shorten the encoding, and most of the drop comes early in $\beta$. At fixed $n$ and $p$,
$\Tenc(\beta)/\Tenc(0)$ against the number of rewired bonds, $2n\beta$, falls on one curve for every size until it levels
off near $\beta=1$ [Fig.~\ref{fig:tencbeta}(a)]: the onset is set by how many bonds are rewired, not by the share
$\beta$. Over $n=64$--512 and $p=0.01$--0.10 (171 pairs: the nine sizes and 19 rates), $\Tenc$ drops by 10\% after
$2n\beta=1.7$--5.4 rewired bonds and by 20\% after 4.9--13.6 (interpolated in $\ln 2n\beta$ between the grid angles),
with no drift in $n$. The graph changes on the same scale: on the graphs of Fig.~\ref{fig:graphs} at $n=64$--256 and
512, $\lambda_2$ doubles after 3.9--6.5 rewired bonds and the mean path length halves after 6.0--14.8. The share of the
drop made by $\beta$, $\Delta\Tenc(\beta)/\Delta\Tenc(1)$ with $\Delta\Tenc(\beta)=\Tenc(0)-\Tenc(\beta)$, has a median
of 0.77 by $\beta=0.099$ ($\theta_{10}$) (quartiles 0.70 and 0.81; 0.56--0.89 over the 171 pairs)
[Fig.~\ref{fig:tencbeta}(b)]. The level reached at $\beta=1$ falls with $n$, from a median over the rates of 0.66 at $n=64$ to 0.33 at $n=512$.

\section{Finite-size scaling and the test at \texorpdfstring{$n=1024$}{n = 1024}}\label{sec:fss}

This section derives Eq.~(\ref{eq:fss}) of the main text and tests it on $n=1024$; Sec.~\ref{sec:exp} gives $p_c$,
$\nu$, and $z$.

\subsection{The standard ansatz}\label{sec:fss-usual}

The observable is $y=\ln\med\,\tau_1$, the logarithm of the median of the first layer $\tau_1$ at which at most one
logical qubit remains [Eq.~(\ref{eq:tau})], one value, or cell, per size, rate, and angle, with error $\sigma$, its
standard deviation over resamples of the graphs.
Near a continuous transition a single relevant field, $\delta=p-p_c$, grows under coarse-graining, and finite-size
scaling~\cite{fisherbarber1972,cardy1996} assumes that the size enters only through its ratio to the correlation length,
$\xi\sim|\delta|^{-\nu}$. Coarse-graining by a factor $s$ maps
\begin{equation}
n\to n/s,\qquad \delta\to s^{1/\nu}\delta,\qquad \tau_1\to s^{-z}\tau_1 ,
\label{eq:rg}
\end{equation}
with $z$ the dynamical exponent. The number of logical qubits is an entropy, of scaling dimension zero, so the time for it
to fall to a fixed level carries only the factor $s^{-z}$~\cite{gullans2020prx}. Choosing $s=n$,
\begin{equation}
\med\,\tau_1=n^{z}\,\Psi\bigl(\delta\,n^{1/\nu}\bigr)
\qquad\Longleftrightarrow\qquad
y=z\ln n+F(x),\quad x=\delta\,n^{1/\nu},\quad F=\ln\Psi .
\label{eq:m0}
\end{equation}
On the ring $n$ is the length of a $(1+1)$-dimensional circuit, and $z=1$ at a conformally invariant critical
point~\cite{gullans2020prl,li2021conformal}. Rewired graphs have no length: their diameter grows only as $\ln n$
(Sec.~\ref{sec:graphs}). We keep the same form with the number of qubits as the size, as for systems without a length
scale~\cite{botet1982}, transitions on small-world networks~\cite{hong2002}, and monitored circuits with sparse nonlocal
gates~\cite{hashizume2022}; $\nu$ then sets how the critical window narrows, $\Delta p\sim n^{-1/\nu}$, and is not a
correlation-length exponent.

At our sizes Eq.~(\ref{eq:m0}) fails. Fitted to $n\le256$, it predicts the medians at $n=512$ with a mean $\chi^2$ of
61.2 per cell, against 3.0 for a correct model; at $\beta=1$ its $p_c$ moves by $+0.018$ and its
$\nu$ by $+0.29$ between $n\le256$ and $n\le512$. Such drifts, common in monitored circuits at accessible
sizes~\cite{zabalo2020,sierant2022,hashizume2022}, have two causes: irrelevant fields add terms that decay as a power of the size~\cite{wegner1972,cardy1996}, and the
scaling field is an analytic function of $\delta$ rather than $\delta$ itself. We include both, following Slevin and
Ohtsuki~\cite{slevin1999,slevin2014}.

\subsection{The modified ansatz}\label{sec:fss-corr}

With a relevant field $u_r(\delta)$, of exponent $1/\nu$, and the leading irrelevant field $u_i(\delta)$, of exponent
$-\omega<0$, Eq.~(\ref{eq:rg}) gives
\begin{equation}
y=z\ln n+\mathcal F\bigl(u_r(\delta)\,n^{1/\nu},\ u_i(\delta)\,n^{-\omega}\bigr).
\label{eq:so}
\end{equation}
Three expansions make Eq.~(\ref{eq:so}) a fit~\cite{slevin1999}. At finite $n$ there is no phase transition, so
$\mathcal F$ is smooth in both arguments; for an irrelevant field that is not dangerous, the second is small at large
$n$, and to first order,
\begin{equation}
y=z\ln n+F(x)+n^{-\omega}G(x),\qquad x=u_r(\delta)\,n^{1/\nu},
\label{eq:so1}
\end{equation}
with $F(x)=\mathcal F(x,0)$ and $G(x)=u_i(0)\,\partial_2\mathcal F(x,0)$, the field taken at its lowest order. The
scaling fields are analytic in $\delta$, and $u_r$ vanishes at $p_c$, so
\begin{equation}
u_r(\delta)=\delta+b\,\delta^2+O(\delta^3),
\label{eq:ur}
\end{equation}
with the linear coefficient set to 1 because $F$ carries the scale. Over the finite range of $x$ that the data cover,
$F$ and $G$ are smooth and are approximated by polynomials of degrees $K$ and $K_G$, written as Chebyshev series,
\begin{equation}
F=\sum_{i=0}^{K}f_i\,T_i(\hat x),\qquad G=\sum_{l=0}^{K_G}g_l\,T_l(\hat x),\qquad \hat x=\frac{2(x-x_{\min})}{x_{\max}-x_{\min}}-1,
\label{eq:cheb}
\end{equation}
with $x_{\min}$ and $x_{\max}$ the extremes of $x$ over the fitted cells. Equations~(\ref{eq:so1})--(\ref{eq:cheb}), with
Eq.~(\ref{eq:ur}) truncated at $\delta^2$, are Eq.~(\ref{eq:fss}) of the main text:
\begin{equation}
y=z\ln n+\sum_{i=0}^{K}f_i\,T_i(\hat x)+n^{-\omega}\sum_{l=0}^{K_G}g_l\,T_l(\hat x),\qquad
x=(\delta+b\,\delta^2)\,n^{1/\nu}.
\label{eq:fitform}
\end{equation}
It has $K+K_G+3$ linear parameters, $z$, the $f_i$, and the $g_l$, and four nonlinear ones, $p_c$, $\nu$, $\omega$, and
$b$: 17 at $K=8$ and $K_G=2$ (Sec.~\ref{sec:fss-cal}). They minimize $\chi^2=\sum_{\text{cells}}[(y-\hat y)/\sigma]^2$, with $\hat y$ the value of Eq.~(\ref{eq:fitform}). The model is linear at fixed $(p_c,\nu,\omega,b)$, so $\chi^2$ is minimized over these four alone, from many starting
points~\cite{golub1973}. The cells at one $n$ share graphs and circuits, so $\chi^2$ is descriptive; forms are compared by
how well they predict a size they were not fitted to, Eq.~(\ref{eq:score}).

\subsection{The cells}\label{sec:fss-data}

There is one cell per size $n=64,96,\ldots,256$, 512, and 1024, rate in each angle's window around $p_c$
(Table~\ref{tab:runs}), and angle. At $n\le256$ each angle's median also uses the graphs simulated at other angles: by
Eq.~(\ref{eq:decomp}), a graph with $m$ rewiring attempts has the same law at every $\beta$, so graph $g$, with $m_g$
attempts, enters the median at $\beta_j$ with the weight
\begin{equation}
w_g=\frac{1}{C_{m_g}}\binom{2n}{m_g}\beta_j^{m_g}(1-\beta_j)^{2n-m_g},
\label{eq:pool}
\end{equation}
with $C_m$ the number of pooled graphs with $m$ attempts; a cell is kept only if the strata $m$ with no pooled graph
carry at most 1\% of the probability. At
$n=512$ and 1024 each angle uses its own 1000 graphs.

A cell enters the fit when $\med\,\tau_1\ge4$ and its median is defined, that is, reached by half the circuits by
$T=16n$; there is no other cut. The final fit has 210 cells on the ring and 454 at $\beta=1$.

\subsection{The orders, on the ring and by prediction}\label{sec:fss-cal}

The ring, whose answer is known, $z=1$~\cite{gullans2020prl,li2021conformal} and
$\nu=1.28(2)$~\cite{gullans2020prx}, checks the form but does not choose it (Table~\ref{tab:ring}, $n\le512$, $K=8$).
Only Eq.~(\ref{eq:so1}) with the quadratic term and $K_G=2$ recovers both, $z=0.994$ and $\nu=1.286$ ($\Delta\chi^2=0.21$
for fixing $z=1$), and it is stable in $K$: $K=8$, 10, 12, and 14 give $z=0.979$--0.994 and $\nu=1.273$--1.302, so we
take $K=8$. A more flexible $F$ also lets the standard ansatz reach these values: at $K=10$, 12, and 14,
Eq.~(\ref{eq:m0}) gives $z=1.015$, 1.014, and 1.018 and $\nu=1.293$--1.301, and fixing $z=1$ costs only
$\Delta\chi^2=2.95$--5.20, while Eq.~(\ref{eq:so1}) without the quadratic term stays at $z=0.82$--0.95. With the
correction $G$, the ring thus rejects the linear field $u_r=\delta$ but cannot separate Eq.~(\ref{eq:m0}) from
Eq.~(\ref{eq:fitform}); the prediction of a larger size does.

The order of Eq.~(\ref{eq:ur}) is chosen by how well a fit to $n\le256$ predicts the $N$ cells at $n=512$,
\begin{equation}
\bar\chi^2=\frac1N\sum_{\text{cells at }n=512}\left[\frac{y-\hat y}{\sigma}\right]^2.
\label{eq:score}
\end{equation}
A correct model scores 3.00(38), not 1, since the cells at one $n$ share circuits and the prediction carries the fit's
uncertainty (synthetic data drawn from the fitted model). Over all angles the linear field scores
5.12 and the quadratic 3.18. On the cells all three forms predict ($\beta\ge0.16$), the cubic form scores 5.66 against
3.00 although it fits better, so we stop at $\delta^2$.

Keeping only the central 70\% of the cells in $x$ gives $\nu=4.88$ and $z=0.26$ on the ring at $n\le512$, so we do not
use it; refitted to $n\le1024$, it moves $p_c(\beta=1)$ by $+0.015$.

\begin{table}[tbp]
\caption{The ring as the calibration: each form fitted to the ring's cells at $n=64$--512 with $K=8$, $z$ free. Known:
$z=1$, $\nu=1.28(2)$. $\Delta\chi^2$: the cost of fixing $z=1$. $^{a}$The fit stops at the bound $\omega=0.1$ of the
search.}
\label{tab:ring}
\begin{ruledtabular}
\begin{tabular}{llllll}
form & $u_r(\delta)$, $G$ & $p_c$ & $\nu$ & $z$ & $\Delta\chi^2(z=1)$ \\ \hline
Eq.~(\ref{eq:m0}) & $\delta$, none & 0.14523 & 1.378 & 0.968 & 23.85 \\
Eq.~(\ref{eq:so1}) & $\delta$, $K_G=1$ & 0.14629 & 1.424 & 0.821 & 112.34 \\
Eq.~(\ref{eq:so1}) & $\delta$, $K_G=2$ & 0.14662 & 1.433 & $0.328^{a}$ & 99.86 \\
Eq.~(\ref{eq:so1}) & $\delta+b\delta^2$, $K_G=1$ & 0.14550 & 1.297 & 0.884 & 32.67 \\
Eq.~(\ref{eq:so1}) & $\delta+b\delta^2$, $K_G=2$ & 0.14475 & 1.286 & 0.994 & 0.21 \\
\end{tabular}
\end{ruledtabular}
\end{table}

\subsection{The test at \texorpdfstring{$n=1024$}{n = 1024}}\label{sec:fss-test}

With every choice fixed, the predictions of each form, fitted to $n\le512$, were stored, with time stamps kept in the
release, before any $n=1024$ cell existed. Over all 50 angles the quadratic form predicts the
1272 medians at $n=1024$ with $\bar\chi^2=3.77$, against 7.91 for the linear form and 7.50 for the cubic; at
$\beta\ge0.16$ it scores 3.34, at the null of the $n\le256\to512$ test, and on the ring 2.75 against 20.43. The prediction fails
only near the ring: at $\theta_1$--$\theta_3$ ($\beta\le0.0092$) every form misses, by 11.96--13.62, and at
$\theta_4$--$\theta_{12}$ the quadratic form scores 4.36. The test also excludes the earlier estimate of a critical point
of $\beta=1$ near $p=0.34$: registered as a power law $y=A+z\ln n$ at fixed $p$, it misses its twelve cells by 89.5,
against 1.12 for Eq.~(\ref{eq:so1}). On the ring the test scored 11 registered cells; the rates $p=0.1425$ and 0.1435
were added afterwards and enter only the final fit.

\section{Critical exponents}\label{sec:exp}

\begin{figure*}[tbp]
\centering
\includegraphics[width=\textwidth]{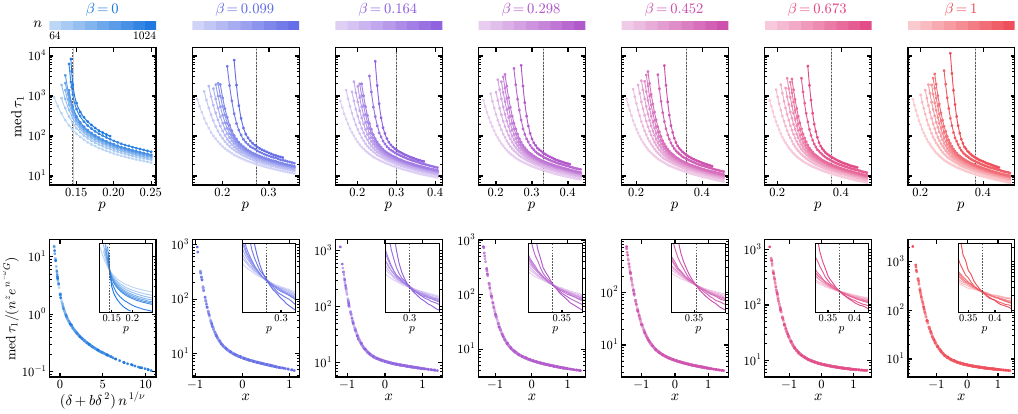}
\caption{Equation~(\ref{eq:fitform}) fitted to $y=\ln\med\,\tau_1$ at $n=64$--1024 at the seven angles of
Fig.~\ref{fig:codes} of the main text, in its colors, the sizes from light ($n=64$) to full (1024). Top: $\med\,\tau_1$ against $p$, with the fit at
each size (lines) and $p_c$ (dashed). Bottom: the collapse, $\med\,\tau_1/(n^{z}e^{n^{-\omega}G})$ against
$x=(\delta+b\,\delta^2)\,n^{1/\nu}$, as in Fig.~\ref{fig:fss}(c), (d) of the main text; insets: the same quantity
against $p$ near $p_c$, where the sizes cross. Error bars: one standard deviation of $y$ over 2000 graph resamples,
smaller than the points.}
\label{fig:fssrow}
\end{figure*}

Equation~(\ref{eq:so1}), with $u_r=\delta+b\,\delta^2$, $K=8$, and $K_G=2$, fitted at each of the 50 angles to
$n=64$--1024, gives $p_c$, $\nu$, and $z$ of Fig.~\ref{fig:fss}. Table~\ref{tab:budget} lists the two ends, and
Fig.~\ref{fig:fssrow} shows the fit at the seven angles of Fig.~\ref{fig:codes} of the main text.

\begin{figure*}[tbp]
\centering
\includegraphics[width=\textwidth]{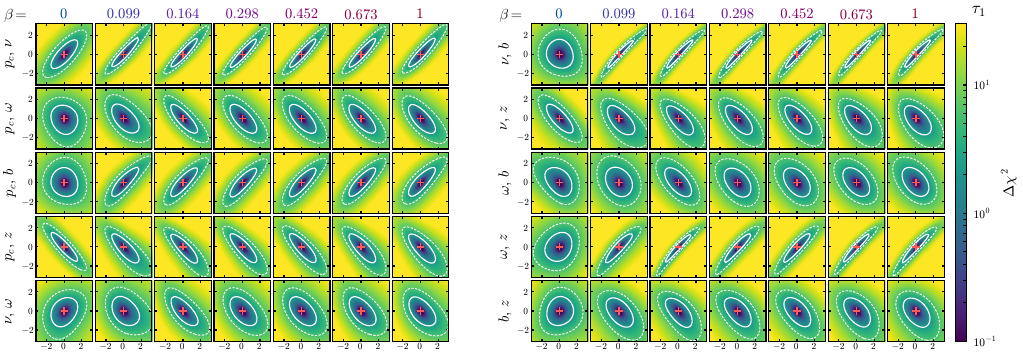}
\caption{Loss landscapes of the fits of Fig.~\ref{fig:fssrow}, at the same seven
angles (columns, $\beta=0$ at the left), for each pair of $p_c$, $\nu$, $\omega$, $b$, and $z$ (rows, in two blocks):
the profile $\Delta\chi^2$ (color, one scale for all panels) with every other parameter refitted, against each
parameter's distance from the best fit (red cross) in units of its half-width $s$ (Table~\ref{tab:land}). Contours:
$\Delta\chi^2=2.30$ (solid) and 6.18 (dashed), the nominal 68\% and 95\% regions of two parameters for independent cells.
Each map is computed on $41\times41$ points over the region where each parameter's own profile stays below
$\Delta\chi^2=12$.}
\label{fig:land}
\end{figure*}

Each of these fits sits at a single minimum (Fig.~\ref{fig:land}): at every angle, the profile $\Delta\chi^2$ of each pair
of $p_c$, $\nu$, $\omega$, $b$, and $z$, with every other parameter refitted, has one closed basin around the fit, with no
second minimum and none at the edge of the search. On the ring $p_c$ trades against $\nu$ and $z$; once the graphs are
rewired, $\nu$ and $b$, $\omega$ and $z$, and $p_c$ and $b$ become strongly correlated. The intervals of
Table~\ref{tab:land} treat the cells as independent; the errors of Table~\ref{tab:budget}, from graph resamples and from
variants of the analysis, are larger.

\begin{table}[tbp]
\caption{The scales of Fig.~\ref{fig:land}: the best fit of each parameter at the seven angles and, in parentheses, the
half-width $s$ of its interval with $\Delta\chi^2\le1$ along its profile, for independent cells (Sec.~\ref{sec:exp}
gives the errors).}
\label{tab:land}
\begin{ruledtabular}
\begin{tabular}{llllll}
$\beta$ & $p_c$ & $\nu$ & $\omega$ & $b$ & $z$ \\ \hline
0 & $0.144736(64)$ & $1.273(11)$ & $1.504(51)$ & $-2.35(13)$ & $0.9946(82)$ \\
0.099 & $0.2729(12)$ & $2.339(68)$ & $0.463(38)$ & $-0.81(23)$ & $0.286(15)$ \\
0.164 & $0.2999(11)$ & $2.297(51)$ & $0.368(34)$ & $-1.14(18)$ & $0.236(19)$ \\
0.298 & $0.3330(11)$ & $2.317(47)$ & $0.452(32)$ & $-1.00(15)$ & $0.270(12)$ \\
0.452 & $0.3534(10)$ & $2.318(40)$ & $0.483(29)$ & $-1.01(12)$ & $0.2782(98)$ \\
0.673 & $0.3707(11)$ & $2.329(42)$ & $0.360(26)$ & $-1.19(13)$ & $0.207(14)$ \\
1 & $0.3759(12)$ & $2.316(45)$ & $0.368(28)$ & $-1.28(15)$ & $0.220(14)$ \\
\end{tabular}
\end{ruledtabular}
\end{table}

On the ring we find
\begin{equation}
p_c=0.1447(6),\qquad \nu=1.27(4),\qquad z=0.99(8)\qquad(\beta=0),
\label{eq:pc-ring}
\end{equation}
with $\omega=1.50$ (16--84\%: 1.43--1.58); fixing $z=1$ costs $\Delta\chi^2=0.42$ ($\chi^2=193.1$ on 193 degrees of
freedom). Both exponents agree with $z=1$ and $\nu=1.28(2)$~\cite{gullans2020prx,li2021conformal}. Since the ring set $K$
(Sec.~\ref{sec:fss-cal}), this is a consistency check; the independent test is the prediction of $n=1024$
(Sec.~\ref{sec:fss-test}).

On fully rewired graphs we find
\begin{equation}
p_c=0.376(5),\qquad \nu=2.3(2),\qquad z=0.22(6)\qquad(\beta=1).
\label{eq:b1}
\end{equation}
Rewiring raises $p_c$ by a factor 2.60, half of it by $\beta=0.081$ [Fig.~\ref{fig:fss}(a)]. At every angle with
$\beta\ge0.16$, $\nu=2.24$--2.47, $z=0.184$--0.278, and $\omega=0.273$--0.483; at $\beta=1$, $\omega=0.368$ (16--84\%:
0.320--0.415). The correction decays slowly, so $z$ and $n^{-\omega}G$ can trade off (Fig.~\ref{fig:land}). The value of $z$ at
$\beta=1$ rests on the form of Eq.~(\ref{eq:so1}): at $p_c$, $\med\,\tau_1$ over $n=64$--1024 is also described by
$(\ln n)^{1.83}$ ($\chi^2/\mathrm{dof}=1.8$), better than by a pure power of $n$ (6.7), and Eq.~(\ref{eq:so1}) assigns the
difference to the correction $n^{-\omega}G$.

\begin{figure*}[tbp]
\centering
\includegraphics[width=\textwidth]{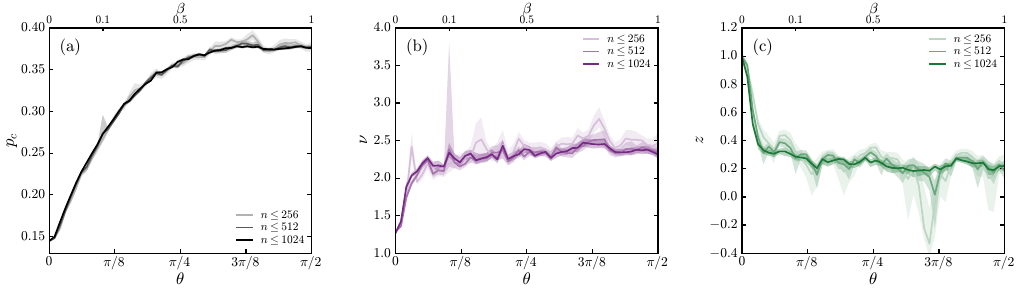}
\caption{The fit of Eq.~(\ref{eq:fitform}) to $y=\ln\med\,\tau_1$ at each of the 50 angles, with the largest size
$n\le256$ (light), 512, and 1024 (full color): (a)~$p_c$, (b)~$\nu$, and (c)~$z$ against $\theta$ ($\beta$ on top), in
the colors of Fig.~\ref{fig:fss} of the main text. Lines: the fits; bands: 16--84\% of the refits to graph resamples
(200 at $n\le256$ and 512, 2000 at $n\le1024$).}
\label{fig:drift}
\end{figure*}

The exponents change fastest in the small-world crossover (Fig.~\ref{fig:drift}): from $\theta_1$ ($\beta=0.0010$) to
$\theta_{12}$ ($\beta=0.141$), $z$ falls from 0.852 to 0.273 and $\nu$ rises from 1.418 to 2.211. There the fits also
depend most on the largest size: from $n\le512$ to $n\le1024$, $p_c$ and $\nu$ move by a median of 4.1 and 3.2 of their
shift's spread over 50 paired graph resamples at $\theta_1$--$\theta_3$, and by 1.39 and 1.05 at
$\theta_4$--$\theta_{12}$, against 0.11 and $-0.02$ at $\beta\ge0.16$, where the median shifts of $p_c$, $\nu$, and $z$ are all at most 0.15 of
their spread.

Each quoted error combines a statistical and a systematic part,
\begin{equation}
\sigma^2=c^2\sigma_{\mathrm{boot}}^2+\sum_s\Delta_s^2 .
\label{eq:err}
\end{equation}
Here $\sigma_{\mathrm{boot}}$ is the standard deviation over 2000 refits to resamples of the graphs, with one draw shared
by all angles and rates~\cite{efron1979}. These refits underestimate the scatter: on 20 synthetic data sets with a known
answer, refitted at $n\le256$, the refits scatter about the truth by $c=1.18$ times $\sigma_{\mathrm{boot}}$ on the ring,
and by 1.64 ($p_c$) and 1.63 ($\nu$) at $\beta=1$; we scale $\sigma_{\mathrm{boot}}$ by these factors, and use it as is
for $z$ ($c=1$). $\Delta_s$ is the largest shift of the estimate among the variants of source $s$
(Table~\ref{tab:budget}): the sizes (without $n=64$; without $n=1024$), the graphs (each angle's own graphs at
$n\le256$), the degrees ($K=10$, 12, 14), the correction ($K_G=1$), and the rate window (without the 20\% of the cells
with the largest $x$; $|x|\le0.75\max|x|$). One ring variant, the window $|x|\le0.75\max|x|$, stops at the bound
$\omega=0.1$ of the search and is left out; with it, the ring's errors would be $p_c=0.1447(8)$, $\nu=1.27(6)$, and
$z=0.99(35)$. The systematic part is the larger for every quantity.

\begin{table}[tbp]
\caption{The error budget of Eq.~(\ref{eq:err}) at the two ends: the shift $\Delta_s$ of each source, the systematic and
statistical parts, and the total, quoted as one error.}
\label{tab:budget}
\begin{ruledtabular}
\begin{tabular}{lllllll}
 & \multicolumn{3}{c}{ring, $\beta=0$} & \multicolumn{3}{c}{$\beta=1$} \\
source & $p_c$ & $\nu$ & $z$ & $p_c$ & $\nu$ & $z$ \\ \hline
without $n=64$ & 0.0002 & 0.018 & 0.020 & 0.0006 & 0.001 & 0.024 \\
without $n=1024$ & 0.0002 & 0.011 & 0.011 & 0.0016 & 0.035 & 0.036 \\
own graphs at $n\le256$ & $<0.0001$ & 0.001 & 0.003 & 0.0012 & 0.038 & 0.022 \\
$K=10$, 12, 14 & $<0.0001$ & 0.015 & 0.005 & 0.0007 & 0.024 & 0.006 \\
degree $K_G$ of $G$ & 0.0005 & 0.009 & 0.073 & 0.0003 & 0.012 & 0.003 \\
rate window & $<0.0001$ & 0.015 & 0.003 & 0.0039 & 0.211 & 0.027 \\
\hline
systematic & 0.0006 & 0.031 & 0.077 & 0.0045 & 0.219 & 0.056 \\
statistical & 0.0001 & 0.016 & 0.011 & 0.0028 & 0.108 & 0.026 \\
total & 0.0006 & 0.035 & 0.078 & 0.0052 & 0.244 & 0.062 \\
\hline
quoted & $0.1447(6)$ & $1.27(4)$ & $0.99(8)$ & $0.376(5)$ & $2.3(2)$ & $0.22(6)$ \\
\end{tabular}
\end{ruledtabular}
\end{table}

\section{Data availability}\label{sec:data}

Every piece of data produced in this study can be reproduced bit for bit. Each graph, gate, and measurement is drawn
from pseudorandom streams fixed by three master seeds, 2026102017 for the graphs, 2026102029 for the gates, and
2026102039 for the measurements, and by the indices of its condition $(n,\beta,p)$, graph, and circuit. The code that
generates, analyzes, and plots the data, with everything else needed to reproduce the results exactly, is openly
available in the folder \texttt{papers/SmallWorldMIPT} of the SparseGF2 repository on GitHub~\cite{sparsegf2}.

\end{document}